\documentclass[useAMS,usenatbib]{mn2e}
\usepackage{graphics,epsfig,psfig}
\usepackage[normalem]{ulem}
\usepackage{xcolor}
\usepackage[]{inputenc,amssymb}

\def \be{\begin{equation}}
\def \ee{\end{equation}}
\def \bea{\begin{eqnarray}}
\def \eea{\end{eqnarray}}
\def\etal{{et al.\ }}

\definecolor{webgreen}{rgb}{0,.5,0}
\definecolor{webbrown}{rgb}{.6,0,0}
\usepackage[pdfpagelabels]{hyperref}
\hypersetup{%
   colorlinks=true,hyperfootnotes=false,%
   breaklinks=true,%
   plainpages=false, bookmarksnumbered, bookmarksopen=true,
   bookmarksopenlevel=1,%
   urlcolor=webbrown, linkcolor=webbrown, citecolor=webgreen,
   }

\title[Escape fraction from OB associations in disk galaxies]{
Narrow escape: how ionizing photons escape from disc galaxies}

\author[Arpita Roy, Biman B. Nath, Prateek Sharma]
{Arpita Roy$^{1,2}$ \thanks{arpita@rri.res.in}, Biman B. Nath$^1$, Prateek Sharma$^2$\\
$^1$Raman Research Institute, Sadashiva Nagar, Bangalore 560080, India\\
$^2$Joint Astronomy Programme and Department of Physics, Indian Institute of Science, Bangalore 560012, India\\
}

\begin{document}

\maketitle

\label{firstpage}

\begin{abstract}

In this paper we calculate the escape fraction ($f_{\rm esc}$) of ionizing photons from starburst galaxies. Using 
2-D axisymmetric hydrodynamic  simulations, we study superbubbles created by overlapping supernovae in OB associations. We calculate the escape fraction of ionizing photons from 
the center of the disk along different angles through the superbubble and the gas disk. 
After convolving with the luminosity function of OB associations, we show that
the ionizing photons escape within a cone  of $\sim 40 ^\circ$, consistent with observations of nearby galaxies. The evolution of the
escape fraction with time shows that it falls initially as cold gas is accumulated in a dense shell. After the shell crosses a few scale 
heights and fragments, the escape fraction through the polar regions rises again. The angle-averaged escape fraction cannot exceed 
$\sim [1- \cos (1 \, {\rm radian})] = 0.5$ from geometrical considerations (using the emission cone opening angle).
We calculate the dependence of the time- and angle-averaged 
escape fraction on the mid-plane disk gas density (in the range $n_0=0.15-50$ cm $^{-3}$) and the disk scale height (between $z_0=10-600$ pc).
We find that the escape fraction is related to the disk parameters (the mid-plane disk density and scale height) roughly
so that $f_{\rm esc}^\alpha n_0^2 z_0^3$ (with $\alpha\approx 2.2$) is a constant. For disks with a given WNM temperature, 
massive disks have lower escape fraction than low mass galaxies. For Milky Way ISM parameters, we find
$f_{\rm esc}\sim 5\%$, and it increases to $\approx 10\%$ for a galaxy ten times less massive.
We discuss the possible effects of clumpiness of the ISM on the estimate of the escape fraction and
the implications of our results for the reionization of the universe.
\end{abstract}

\begin{keywords} 
(ISM:) H ii regions -- (galaxies:) intergalactic medium -- galaxies: starburst -- ISM: bubbles   
\end{keywords}

\section{Introduction}
The evolution of galaxies is tuned by various regulatory mechanisms. Galaxies eject gas, radiate photons,
accelerate high energy particles, thereby affecting their surroundings, which in turn influence their evolution. The
leakage of ionizing photons from galaxies is one such crucial mechanism, since it produces an ultraviolet background radiation
that significantly influences the evolution of galaxies in the universe. The sources of the ionizing photons determine the spectrum
of the background radiation in the intergalactic medium (IGM), which has important implications, e.g., in the reionization
of the universe at $z \ge 6$. However, it remains uncertain whether star forming galaxies or active galactic nuceli (AGNs) were the main 
contributors to the epoch of re-ionization. 
One of the main reasons for the uncertainty is our lack of knowledge of how the local ionizing background in individual galaxies is produced, 
because all extrapolations to high redshift depends on it.

The understanding of this background radiation depends on the knowledge of $f_{\rm esc}$, the fraction of ionizing photons from
massive stars that can escape the galaxies. Several workers have estimated this `escape fraction' from both theoretical considerations 
and observations. Since there is a wide disagreement in the value of escape fraction, we summarise below these estimates and 
the methods used to produce them.

\noindent
{\it Direct observations}: Observational estimates of the escape fraction of ionizing photons from extragalactic objects became 
possible with the new generation of UV telescopes in 1990s. \citet{leitherer1995}
observed the luminosity of four starburst galaxies at $900 \, \AA$, and estimated $f_{\rm esc} \le 3 \%$. This estimate was revised upwards 
by \citet{hurwitz1997} to lie between $3\hbox{--}57 \%$ after taking into account a detailed model of interstellar absorption.
 \citet{bland1999} used 
H$\alpha$ measurements of the Magellanic Stream to infer $f_{\rm esc}\le 6\%$ for our Galaxy.
Observations of CII interstellar absorption line at $\lambda 1036$ with {\it FUSE} allowed \citet{heckman2001}  to estimate $f_{\rm esc} \le 6\%$ for a set of five bright starburst galaxies. 
 While comparing their estimate with
that from starburst galaxies with outflows, \citet{heckman2001} posed the question if the outflowing gas affect $f_{\rm esc}$ and concluded that
outflows do not necessarily increase its value by creating additional channels in the ISM. However, in their later study with a bigger sample, 
\citet{heckman2011} concluded that outflows from starburst can significantly increase the escape fraction.
They also discovered four objects, whose central regions are very compact ($\sim 10^2$ pc) and massive ($\sim 10^9$ M$_\odot$), and which 
had significant outflows. They have recently shown that one of these extreme objects has an escape
fraction $\sim 20\%$ (\citealt{borthakur2014}). Working with {\it FUSE} archival data, \citet{leitet2013} estimated $f_{\rm esc} \le 2.5 \%$ for 
a local starburst. Recently, \citet{zastrow2013} obtained the emission-line ratio maps of [SIII]/[SII] for a few dwarf starbursts, and observed 
the ionizing photons to be confined within a cone of opening angle $\sim 40^\circ$. This means that the direct measurement of the escape fraction
by modeling the absorption in UV lines depends sensitively on the orientation. They also pointed out that the escape fraction is large for 
galaxies  older than the time needed by supernovae to create
pathways for ionizing photons  ($\sim 3$ Myr), but young enough so that O stars (the main contributors to ionizing photons)
are still present (main sequence life time of $\sim 5$ Myr).

Observational constraints on the escape of ionizing photon from high redshift galaxies are sparse. Soon after the Lyman Break Galaxies (LBG) 
were discovered at $z\sim 3$, the escape fraction from them were estimated to be $10\hbox{--}20\%$ (\citealt{adelberger2000}; although 
\citealt{haehnelt2001} argued that it could be as large as $50\%$). Recent observations by \citet{nestor2011} have yielded $f_{\rm esc} \sim 10\%$, 
and
\citet{cookej2014} found $f_{\rm esc}\sim 16\pm 4 \%$ for LBGs. At a lower redshift, $z \sim 1$, observations of starbursts in the 
GOODS field have suggested $f_{\rm esc}\le 2\%$ (\citealt{siana2010}).

\noindent
{\it Constraint from UV background radiation:} It is possible to put constraints on $f_{\rm esc}$ from the observations of the UV 
background radiation and considering the possible contribution from starburst galaxies (or AGNs). \citet{shull1999} argued that starbursts 
would be able to contribute as much to the UV background radiation as AGNs if $f_{\rm esc} \ge 5\%$. \citet{inoue2006} reconsidered this issue 
in light of more recent observations, and concluded that $f_{\rm esc}\le 1\%$ at $z\le 1$ and increases to $\sim 10\%$  at $z\ge 4$. 
\citet{fujita2003} simulated superbubbles in dwarf starburst galaxies, and argued that high redshift galaxies may make a significant 
contribution to the background radiation if the escape fraction is $\ge 20\%$.  Recently, \citet{kollmeier2014} have
suggested that the intensity of the UV background radiation at low redshift ($z<0.4$) is likely to be higher than previously thought, which
would imply a corresponding increase in the required escape fraction. However, the assumed fiducial value of
escape fraction at low redshift is quite low in this calculation, the `minimal reionization model' of 
$f_{\rm esc}=1.8 \times 10^{-4} (1+z)^{3.4}$ (\citealt{haardt2012}).

\noindent
{\it Constraint from the epoch of reionization:} It is also possible to put constraints on the escape fraction from the requirement of 
explaining the observations related to reionization (the minimum redshift by which the universe is believed to have been fully ionized and 
the total Thomson optical depth). \citet{madau1996} estimated the required $f_{\rm esc} \sim 50\%$ in order to keep the universe ionized at 
$z\sim 5$. However, this calculation depends on the assumption of luminosity function of galaxies at high redshift or, if a theoretical mass 
function is used, on the assumption of star formation efficiency in galaxies. \citet{inoue2006} estimated the minimum requirement for 
reionization, as $f_{\rm esc}=1.8 \times 10^{-4} (1+z)^{3.4}$, which implies $f_{\rm esc}\sim 2\%$ at $z=3$. Recently \citet{mitra2013} have 
used their semi-analytical model of galaxy formation to compare with the observed luminosity functions at high redshift, and then to 
calculate the requirement from reionization observations. They concluded that reionization requires $f_{\rm esc} \sim 7 ( \pm 5) \%$ 
at $z=6$, and a mild (but uncertain) increase to $\sim 18^{+33} _{-13} \%$ at $z=8$.

\noindent
{\it Theoretical calculations}: \citet{dove1994} analytically calculated the escape fraction for the Milky Way ($\sim 10\%$), 
considering HII regions around OB association that likely produces `HII chimneys' for ionizing photons to escape. They argued that this 
process could explain the ionizing radiation needed for the existence of the Reynolds layer of warm ionized medium.  \citet{wood2000} 
estimated the escape fraction considering a disk galaxy in steady state, with sources embedded in it, using a 3-D radiation transfer code. 
They used a stratified disk with redshift dependent parameters, using the prescriptions of \citet{mo1998}. Naturally, their disks had large 
density and were thin at high redshift, and their escape fraction rapidly decreased with an increasing redshift. Galaxies with mass 
$10^{12}$ M$_\odot$ at $z=0$ have $f_{\rm esc}\sim 1\%$, whereas smaller galaxies, say with $10^9$ M$_\odot$, have similar escape fraction at
 $z \sim 3$. On average they predicted an escape fraction $\le 1\%$ at $z\sim 10$. A similar calculation by 
\citet{benson2013} took into account the effect of high energy photons in the case of sources of hard spectrum. Instead of radiation transfer 
through a disk, \citet{ferrara2013} considered the evolution of ionization fronts in dark matter halos containing gas at $10^4$ K, and found 
the escape fraction to increase with redshift during reionization.

In contrast to these calculations, \citet{dove2000} and \citet{clarke2002} emphasised the importance of the time evolution of superbubbles
in this regard. It is clear that the structure of the ISM must
be important in determining the leakage of ionizing photons.
The typical HI column density in galaxies ($\ge 10^{21}$ cm$^{-2}$) is much larger than needed to make the ISM on 
average opaque to these photons, since only $N_{HI} \ge 10^{17}$ cm$^{-2}$ is needed to shield them. 
\citet{dove2000} considered the OB associations in the Milky Way and propagation of ionizing photons through the superbubbles triggered by them. These superbubbles first decelerate during their early evolution, following the results of \citet{weaver1977} for the self-similar
evolution of stellar wind driven bubbles. However, after they reach the scale height of the stratified disk, they accelerate 
(\citealt{kompaneets1960, roy2013}),
which can fragment the superbubble shells through Rayleigh-Taylor instability and create pathways for ionizing photons to escape.
\citet{dove2000} estimated the escape fraction to be $\sim 6\hbox{--}15 \%$, depending on the assumption of star formation history. This idea was carried forward with the help of hydrodynamic simulations by \citet{fujita2003}, who
 studied the time evolution
of the escape fraction for OB associations of a given mass (with number of OB stars $N_{O}=40000$), and stressed that superbubbles can 
effectively trap the ionizing photons before blowout. 
Their work explained the time evolution of $f_{\rm esc}$ in light of the two competing processes, between a diminishing source of 
ionizing photons from an evolving OB association and the increasing number of pathways created by the superbubbles. Therefore, the observed 
low values of $f_{\rm esc} \ge 5\%$ in 
local dwarf galaxies could be explained if they were young (before the feedback processes could carve out enough channels in the ISM);  
the time-averaged $f_{\rm esc}$ could be larger than these observational estimates. They estimated the escape fraction in high redshift 
disk galaxies to be $\sim 20\%$ and highly dependent on the dynamics of superbubbles in them (because at high density, escape of ionizing 
photons would be difficult without the aid of superbubbles).

Several authors have also used hydrodynamic  simulations for galaxy formation, with star formation and their feedback processes included in 
them, and estimated the escape fraction. These simulations have less control on the parameters and therefore the results are often difficult 
to interpret. \citet{gnedin2008} found the escape fraction to decrease rapidly with a decreasing galactic mass, because of the increasing disk 
thickness and the paucity of young stars on the periphery of the disks. On the contrary, \citet{razoumov2010}  estimated the escape fraction 
to increase with a decreasing galactic mass and  an increasing redshift (see also \cite{wise2014}), reaching an average of $\sim 80\%$ at $z\sim 10$. \citet{yajima2011} 
concluded that massive disks were more clumpy than low mass disks, and this decreased $f_{\rm esc}$ in large galaxies, since stars were 
embedded deep in the clumps. They estimated $f_{\rm esc}\sim 40\%$ for $10^9$ M$_\odot$ galaxies  and $\sim 7\%$ for $10^{11}$ M$_\odot$ 
galaxies.
\citet{paardekooper2013} and \citet{hutter2014}  also arrived at similar conclusions. However, this view of the effect of clumps on the escape fraction runs opposite to that of 
\citet{fernandez2011}, who argued that fewer, high-density clumps would lead to a greater escape fraction than in the case of more numerous 
low-density clumps.
\citet{kimm2014} found from the study of a large number of galaxy halos in their simulation that on average $f_{\rm esc}\sim 10\%$ with little 
variation with galactic mass and redshift, although instantaneous values of the escape fraction could reach $\ge 20\%$. They also found that 
run away OB stars could increase the average escape fraction to $\sim 14\%$.

Our goal in this paper is to extend the previous works on the escape fraction by focusing on the effect of superbubbles. The distinctive 
features of this work are:
(1) controlled hydrodynamic numerical experiments on a wide range of disc parameters, with disc densities ranging between $0.5\hbox{--}50$ cm$^{-3}$ 
and scale heights between $10\hbox{--}600$ pc; 
(2) the escape fraction is weighted by the luminosity function of OB associations 
and does not use OB associations of a particular size. Our strategy is to focus on the effect of important disc parameters on the escape fraction, 
rather than to explore a large number of effects at once, such as ISM clumpiness 
or cosmological mergers. 
In our simulations we use gas without any initial clumpiness, but we discuss the effect of clumpiness on the escape fraction by considering the 
covering fraction of the fragmented superbubbles.

The paper is organized as follows. In section \ref{sec:setup} we present the numerical set up, including the model for superbubbles and the warm disk. 
In section \ref{sec:esc}
we present the method for calculating the escape fraction, which is a function of viewing angle, time and the number of OB stars, in addition to 
the disk parameters ($n_0$, $z_0$). In section \ref{sec:results} we present our 
results on the escape fraction, including variation with redshift and the galactic mass; we also discuss the influence of clumpiness. Section 
\ref{sec:discussion} discusses 
and compares our results with previous works, and section \ref{sec:conc} summarises and concludes our paper.

\section{Numerical setup}
\label{sec:setup}
We begin with a description of our numerical simulation set up.
We use the ZEUS-MP code (Hayes et. al. 2006), a second 
order accurate Eulerian hydrodynamics code. 
We carry out 2-D axisymmetric simulations of
the following governing equations (see also \citealt{roy2013}),
\bea
\label{eq:mass}
&&{d\rho \over dt}=-\rho \nabla {\bf .v} +S_{\rho}(r) \,, \\
&& \rho {{d{\bf v}} \over {dt}}=-\nabla p + \rho \bf{g}\,,\\
&& {de \over dt} =-q^{-}(n,T)+S_{e}(r),
\label{eq:euler_eqns}
\eea
where $\rho$, {\bf v} are the fluid density and velocity respectively, $S_{\rho}$ and $S_e$ 
are the mass and energy source terms respectively as discussed below, $p$ is the thermal pressure of the medium and 
$e =3p/2$ is the internal energy density for a monoatomic gas, 
${\bf g}$ is the disk gravity, $q^{-}=n_en_i\Lambda(T)$ is the energy loss term due to radiative cooling with 
$n_e$ ($n_i$) being the electron (ion) number density, and  
$\Lambda(T)$ is the cooling function. We use the \citet{sutherland1993} cooling function for solar metallicity 
in the temperature range $10^4\hbox{--}10^8$ K; below $10^4$ K $\Lambda(T)$ is set to zero. 
The initial isothermal gas temperature is assumed to be $10^4$ K. All supernovae are assumed to go off
at the center of the disk.

We have used spherical $(r,\theta)$ coordinates for our simulations. Our radial grid extends from
$1$ pc ($r_{\rm {min}}$) to $2$ kpc ($r_{{\rm max}}$); in some higher 
$N_{O}$ runs ($N_{O}\ge 10^4$, $N_O$ being the number of O stars) the outer boundary 
extends up to $r_{\rm max}=3$ kpc. In the highest density cases ($n_0=50$ cm$^{-3}$) the inner radial boundary is
$r_{\rm min}=0.5$ pc, maintaining the strong shock condition (see eqn 4 in \citealt{sharma2014}). The angle $\theta$ runs from 0 to 
$\pi$ and $\phi$ runs from 0 to 2$\pi$. In the radial direction we use logarithmically spaced grid points to 
better resolve the smaller scales. The inner boundary is at 
$r_{\rm {min}}<r_{{\rm in}}$ (energy injection radius; see section 2.2), such that there are equal number of grid points between 
$r_{\rm {min}}$ and 
$(r_{\rm {min}}r_{{\rm max}})^{1/2}$, as there are  between $(r_{\rm {min}}r_{{\rm max}})^{1/2}$ and $r_{\rm {max}}$. 
We use uniformly spaced grid points in $\theta$. We use outflow 
boundary condition at the outer radial boundary. We adopt inflow-outflow boundary condition at the inner radial boundary. 
We apply reflective boundary condition in the $\theta$-direction. 

Our runs with various parameters use different resolutions. Specifically, high density disks (higher $n_0$) with lower energy injection 
(smaller $N_{O}$, the number of O stars) result in extensive formation of multiphase gas, and the number of dense clumps 
increases with an increasing resolution (this is true in all simulations which do not resolve the transition layers between hotter 
and cooler phases; 
e.g., see \citealt{koyama2004}). The low density and higher $N_{O}$ runs do not show much multiphase gas and are less sensitive 
to resolution. Detailed resolution studies are discussed in Appendix A.
Various parameters for our different runs, including resolution, 
are mentioned in Table \ref{table:res}.

\begin{table}
\caption{Parameters for various runs} 
\centering 
\begin{tabular}{|c|c|c|c|c|} 
\hline\hline 
$n_0$ (cm$^{-3}$) & $N_{O}$ & Resolution & $r_{\rm min}$ (pc) & $r_{\rm in}$ (pc) \\ [0.5ex] 
\hline 
0.15, 1, 1.5 & 100 $\textendash 10^5$ & $256\times 128$ & 1 & 2 \\ 
5, 15 & 100 $\textendash$ 300 & $512 \times 512$ & 1 & 2 \\
5, 15 & 600 $\textendash 10^5$ & $256\times 128$ & 1 & 2 \\
50 & 100 $\textendash 10^5$ & $256\times128$ & 0.5 & 1  \\[1ex] 
\hline 
\end{tabular}
\label{table:res} 
\end{table}

The CFL number is the standard value 0.5, but in high density 
( $n_0 \ge 1.5$~cm$^{-3}$) and low $N_{O}$ cases (
$ N_{O} <10^4$) we use 
the CFL number of 0.2 as it is found to be more robust. We have carried out a large number of runs to 
cover a range of values in $n_0$, $z_0$ (disk scale-height) and $N_{O}$. This was necessary to obtain the key result of our paper discussed in section \ref{sec:results}.3.
We have carried out 2-D axisymmetric simulations because 3-D simulations are very expensive due to a larger number of grid points and 
a much smaller stability time step. In Appendix \ref{app:2D3D} we show that the results for our 2-D fiducial run are similar to the 
results obtained in 3-D.

\subsection{The warm neutral disk}

The vertical structure of the thin disk is determined by self-gravity and gas temperature. 
Hydrostatic equilibrium for the gas in $z$-direction is given by, 
\begin{equation}
 {{dp(z)} \over {dz}} = -\rho(z) g(z) 
 \label{eq:hydro_eq}
\end{equation}
where $p$ is the thermal pressure of the gas, $\rho(z)$ is the density and $g(z)$ is 
the vertical disk gravity, the $z-$ component of ${\bf g}$. 
Using Poisson's equation along with eqn \ref{eq:hydro_eq} leads to 
the vertical density distribution of the disk gas (Spitzer 1942),
\begin{equation}
 n(z) = n_0 \, {\rm sech}^2\Bigl ({{z} \over {\sqrt{2}z_0}} \Bigr ) \,, \quad z_0 = { {c_s} \over {\sqrt{4\pi G \mu m_p n_0}} } \,.
 \label{eq:z_distribution}
\end{equation}
Here, $n_0$ is the mid-plane density of the disk ($z=0$) and
$z_0$ is the scale height of the gas in the disk, and $c_s=\sqrt{k_bT/\mu m_p}$ is the isothermal sound speed of the gas. 
 The corresponding value of $g(z)$ is, 
 \begin{equation}
  g(z)={ {\sqrt{2}k_b T} \over {\mu m_p z_0} } \tanh \Bigl ({ z \over {\sqrt{2}z_0} }\Bigr ) \,.
  \label{eq:g}
 \end{equation}
We consider only the self-gravity of the initial stratified gas, and assume it to be constant with time, which is a 
caveat in our calculations. Also, we assume that the initial ISM is non-clumpy.

The equilibrium value of $n_0$, $z_0$ and $T$ are related as (eqn \ref{eq:z_distribution})
\be
\label{eq:HSE}
z_0 = 257~{\rm pc} \left( \frac{T}{10^4 {\rm K}} \right)^{1/2} \left ( \frac{n_0}{0.5~{\rm cm}^{-3}} \right )^{-1/2},
\ee
where we have used $\mu=1.33$.
Later we vary $n_0$ and $z_0$ independently to study the variation of escape fraction as a function of these disk parameters (c.f. Fig. 
\ref{fig:contour_plot}). We note that $c_s$ needs to be adjusted with $(n_0,z_0)$ to obtain a self-consistent hydrostatic equilibrium 
(Eqs. \ref{eq:z_distribution} \& \ref{eq:g}). 
We use the initial disk temperature of $10^4$ K, corresponding to the thermally stable warm neutral phase
(\citealt{wolfire2003}). We, therefore, keep the initial temperature fixed at $10^4$ K, even when we vary $n_0$ and $z_0$ independently 
of each other. This means that our  disks with general $n_0$, $z_0$ parameters are not in perfect hydrostatic balance, except when they satisfy
eqn \ref{eq:HSE}.
However, we note that for all the cases considered here, the dynamical time scale of superbubbles breaking through the disk (eqn \ref{tdyn}), which is important for the determination of 
escape fraction, is always shorter than the gravitational time scale (free-fall) for the disk to evolve. Therefore, the disks
are stable for the time scale of importance in our calculation of escape fraction. 
Most of our runs highlighted in various figures correspond to a self-consistent
hydrostatic equilibrium for $10^4$ K (closely satisfying eqn \ref{eq:HSE}).

We only consider a warm neutral disk at $10^4$ K and do not include a cold neutral (100 K) component. This is justified because 
the cold neutral medium is
expected to be clumpy. Moreover, even if we consider a self-consistent cold disk, the scale height $z_0$ should be much smaller
$\propto (T/n_0)^{1/2}$ 
(eqn \ref{eq:z_distribution}). The fraction of LyC photons absorbed by the disk is roughly given by  (c.f. eqn \ref{eq:esc_frac_1_theta})
$
4\pi \alpha_H^{(2)} n_0^2 z_0^3/S,
$
 which scales as $p^{1/2} T$ ($p$ is pressure). Thus, for a warm and cold disk at pressure equilibrium the escape 
 fraction is dominated by the hotter ($10^4$ K) warm disk.

\subsection{Superbubble implementation}

The mass and energy source functions in Eqs. \ref{eq:mass} \& \ref{eq:euler_eqns} are
  applied in a small enough volume such that radiative losses do not quench the formation of a superbubble (\citealt{sharma2014}).
 The mass source function $S_{\rho}=\dot{M}_{\rm in}/(4\pi r_{\rm in}^3)$,
    where $\dot{M}_{\rm in}=N_{O}M_{\rm ej}/t_{O}=6.33\times 10^{18} N_{O}$ g s$^{-1}$, $M_{\rm ej}$ is
    the ejected mass in a single supernova explosion
     (chosen as 1$M_{\odot}$), $t_{O}$ is 10 Myr (the lifetime of O stars), 
   $N_{O}$ is the number of O stars present {\em initially}, or equivalently, the total number 
   of supernova explosions within time $t_O$. We note that the final results in our simulations are insensitive
      to our choice of $M_{\rm ej}$, which only affects the structure of the hot/dilute gas within the superbubble. The energy source function, which mimics the energy
  input by supernova explosions within $r_{\rm  in}$, is given by $S_e=\mathcal{L}/(4\pi r_{\rm in}^3)$ , where $\mathcal{L}$ is the mechanical 
luminosity of supernovae. We consider $r_{\rm in}=
  2$ pc in all our simulations except for  runs  with $n_0=50$~cm$^{-3}$, 
  for which we take $r_{\rm in}=1$ pc to prevent artificial cooling losses.

We run our simulations for a period of 10 Myr, the average lifetime 
of O9.5 stars (\citealt{chiosi1978,meynet2003,weidner2010}), the least massive O stars of mass $20.8 \rm M _{\odot}$ 
(\citealt{vacca1996}), since O stars 
are thought to produce most of the ionizing
 photons (\citealt{dove1994}). 
 The mechanical luminosity produced by O-stars in the association is,
   \be
  \mathcal{L}=N_{O}E_{\rm ej}/t_{O}=3.16 \times 10^{36}\, N_{O} \, {\rm erg} \, {\rm s}^{-1} \,,
   \label{eq:mechlum}
   \ee
   where $E_{\rm ej}=10^{51}$ erg is the explosion energy of one SN ($10^{51}$ erg). 

\begin{figure}
\centerline{
\epsfxsize=0.52\textwidth
\epsfbox{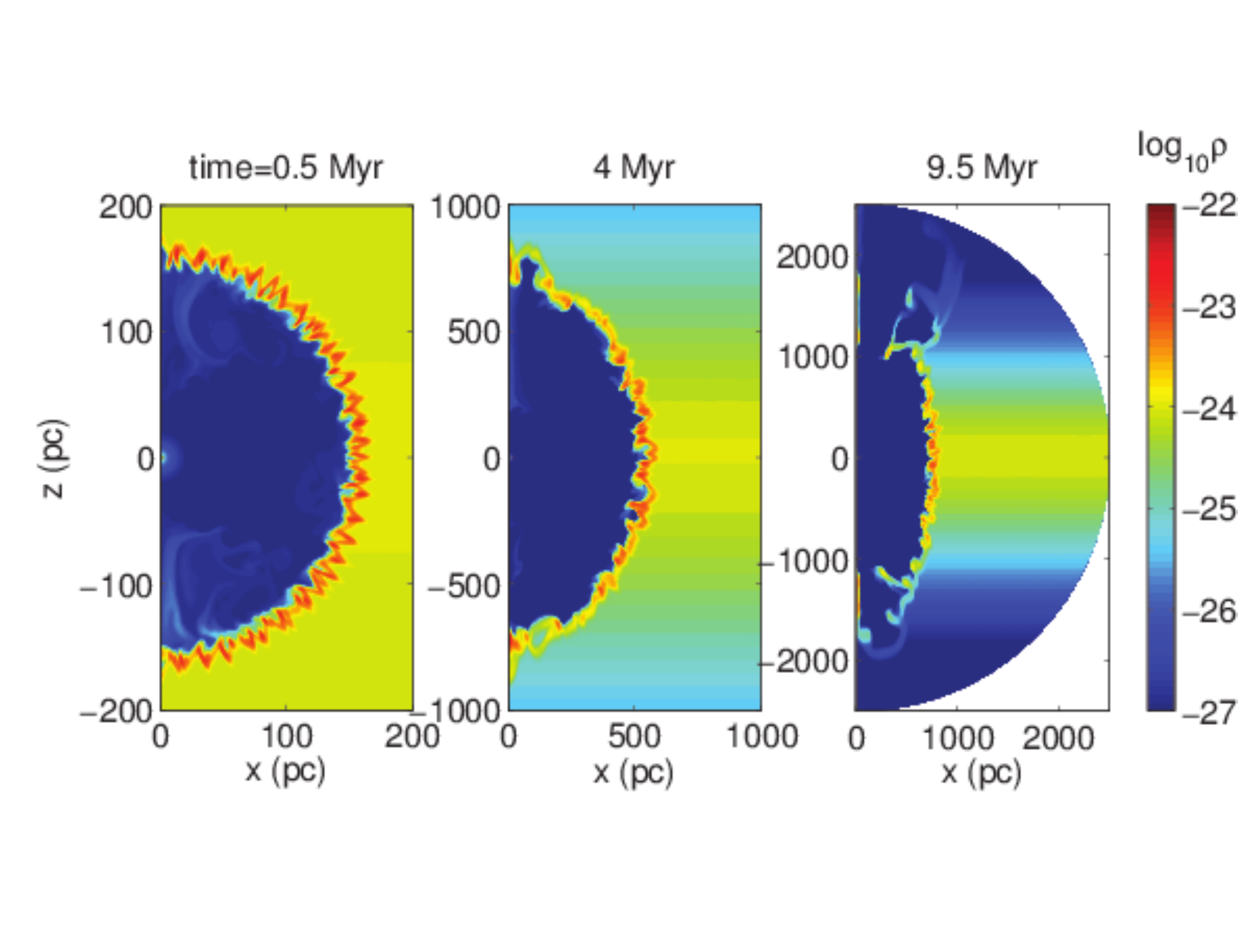}
}
\caption{
Density contour plot of the superbubble at different times (0.5, 
4.0, 9.5 Myr) for $n_0=0.5$ cm$^{-3}$, $z_0=300$ pc and $N_{O}=10^4$. Early, intermediate and late stages of 
superbubble evolution are shown. Notice the low density cone through which 
photons should escape at late times.
 }
\label{fig:den_contour}
\end{figure}

The escape fraction, which we describe shortly in the next section, depends very crucially on the structure of the ISM, in particular to the low 
density channels opened by the expanding superbubble. As studied in detail in \citet{roy2013}, the evolution 
of a superbubble for a sufficiently strong starburst shows two stages: first, the Sedov-Taylor stage when the outer shock radius is smaller than the scale 
height; and second, the fast breakout of the superbubble due to thermal and Rayleigh-Taylor instabilities after it crosses a few scale-heights. The escape 
fraction for a density-bounded (Str\"omgren radius $>$ disk scale height) disk is expected to decrease with time in the first stage as photons are 
absorbed in the dense shell. After breakout, the escape fraction increases with time because of opening of low density channels in the ISM. We see this
effect in the time evolution of the escape fraction described later. Figure \ref{fig:den_contour} shows the density contour plots of our fiducial  run 
($n_0=0.5$ cm$^{-3}$, $z_0=300$ pc, $N_{O}=10^4$) at early,
intermediate and late times. As the escape fraction is intimately connected to the ISM 
porosity, it is useful to remember these various stages of superbubble evolution in order to interpret our results.

\section{Calculation of escape fraction}
\label{sec:esc}
In this section we describe the formalism to calculate the escape fraction assuming photoionization equilibrium (ionization rate 
equals recombination rate). We can calculate the number of photons absorbed per unit time, and hence the number of photons 
escaping along different directions and at different times. Thus, the escape fraction is a function of angle ($\theta$), time ($t$) and 
disk/starburst parameters ($n_0$, $z_0$, $N_{O}$). Since the distribution of OB associations is similar in different regions, 
we average our escape fraction in time, angle, and the number distribution of OB associations.

\subsection{Ionization Equilibrium}
In order to calculate the fraction of ionizing photons that escapes the
disk, consider the ionizing photons emitted within a solid angle $d \Omega$ 
within angles
$\theta$ and $\theta + d\theta$. 
First consider the case of ionization 
equilibrium in the disk gas, and assume that all ionizing
photons are absorbed in the medium, so that the escape fraction is zero. 
If 
$S$ denotes the time-dependent luminosity of ionizing photons (number of ionizing photons produced per unit time 
$\propto N_O$; 
discussed in section 3.2), and $\alpha_H ^{(2)}$ 
denotes the recombination
coefficient for case B (`on the spot' ionization case), then we have in the 
case of ionization equilibrium in a solid angle $d \Omega$,
\begin{equation}
S {d\Omega \over 4 \pi}= \int \alpha_H ^{(2)} n_H ^2 (r) r^2 dr d\Omega \,,
\label{eq:all_ph_abs}
\end{equation}
where $n_H (r) $ is the number density of hydrogen (which is equal to the electron/proton density within the ionized bubble). These considerations 
apply in the standard calculation of Str\"{o}mgren sphere \citep{dyson1997}, which is ionization bounded.

In general, however, all the ionizing
photons will not get absorbed; some will escape and thus eqn \ref{eq:all_ph_abs} will not hold. In this case the ISM is density bounded. Therefore, the escape 
fraction of ionizing photons in an angle between $\theta$
and $\theta + d\theta$ can then be written as,
\begin{eqnarray}
 f_{esc}(\theta,t,N_{O};n_0,z_0)&=&{ {{S d\Omega/4\pi}-{\int_0^{\infty}\alpha_H^{(2)} n_H^2(r)r^2dr \, d\Omega}} \over {S d\Omega/4\pi} }\, \nonumber\\
 &=&1-{{4\pi \alpha_H^{(2)}} \over {S}}\int_0^{\infty} n_H^2(r)r^2dr,
 \label{eq:esc_frac_1_theta} 
\end{eqnarray}
where we have indicated the dependence of the escape fraction on various parameters in parentheses. Later we will average the 
escape fraction over time, angle and number of O stars. The averaged escape fraction is denoted as $\langle f_{esc} \rangle$ where the
parameters over which averaging is done denoted in subscript; e.g., $\langle f_{esc} \rangle_{\theta,t}$ denotes the time- and angle-averaged 
escape fraction as a function of $N_{O}$ for a fixed $n_0$ and $z_0$.
Note that the photon luminosity ($S \propto N_O$) is time dependent, and for a given OB association $S$ decreases abruptly after the most massive 
stars die off (c. f. Fig. \ref{fig:schem_diag}).


The expression in eqn \ref{eq:esc_frac_1_theta}  is valid when the recombination time scale is shorter than other time scales in the problem,
e.g., the dynamical time scale. 
This puts a condition on $n_0$ and $z_0$ (mid-plane density and scale height respectively) for which we can calculate 
the escape fraction using eqn \ref{eq:esc_frac_1_theta}.
The recombination time scale is given by,

\begin{equation}
  t_{\rm reco} \equiv {1 \over {n\alpha_H^{(2)}(T)}} = {1 \over {4n_H\alpha_H^{(2)}(T)}} = 0.04 n_H^{-1}T_4^{3/4} {\rm Myr}\,,
  \label{eq:recomb_time}
 \end{equation}
 where we have written $n=4n_H$, since most of the recombinations occur in the dense shell and $\alpha_H^{(2)}(T)=2 \times 10^{-13} T_4^{-3/4}$ 
$\rm cm^{-3}$ $\rm s^{-1}$ (\citealt{dyson1997}). This means that the balance between ionization and recombination is achieved in a very short 
time.

 The radius of a superbubble in a uniform density medium with density $\rho$ is given by \citep{weaver1977},
 \begin{equation}
  r=\bigl({{\mathcal{L}t^3} \over {\rho}}\bigr)^{1/5}\,.
  \label{eq:pos_lum}
 \end{equation}
 For a scale height $z_0$ of the gas distribution in the disk, the dynamical time is given by (the time taken by
 the superbubble to reach the scale height),
\be
 t_d=\Bigl ({ {\rho_0z_0^5} \over \mathcal{L}} \Bigr ) ^{{1 \over 3}}
 = 4 n_H^{1/3} \, z_{0,100 pc}^{5/3} \, N_{O}^{-1/3}~{\rm Myr},
 \label{tdyn}
\ee
where $\mathcal{L}$ is the mechanical luminosity  (eqn \ref{eq:mechlum}), $n_H$ is hydrogen number density in cm$^{-3}$ and 
$z_{0, 100 pc}=z_0/(100 \, {\rm pc})$, and we have assumed $\mu=1.33$.
The condition $t_{\rm reco}<t_d$ implies,
\begin{equation}
 { {n_H^4z_{0,100 pc}^5} \over {N_{O}} }>10^{-6}\,.
 \label{eq:dyn_time} 
\end{equation}
This means that the assumption of ionization equilibrium holds for the relevant ISM parameters and the use of equation 
\ref{eq:esc_frac_1_theta} is valid. This assumption breaks down only for low mid-plane densities and small scale-heights, 
as discussed in section \ref{sec:results}.


\subsection{Stellar ionizing luminosity}
The escape fraction of ionizing photons depends on the total ionizing 
luminosity $S$ (eqn \ref{eq:esc_frac_1_theta}), which in turn depends 
on the size of the OB association, characterised by $N_{O}$, the 
total number of O stars. The time evolution of the ionizing luminosity depends 
on the initial mass function (IMF) of the OB association and how 
the ionizing luminosity and main sequence life time of stars depend on their
 masses. We use Starburst99 (\citealt{leitherer1999}) to calculate the evolution of the ionizing
 luminosity $S(t)$ for an OB association. The time dependence of LyC luminosity, $S(t)$, is assumed to be the same for all OB associations, 
 but $S(t)$ scales linearly with the number of O-stars ($N_O$). We have used the case
 of an instantaneous starburst and assumed Salpeter
 IMF between $0.1$ and $100$ M$_\odot$. Note that we have only considered O stars in our calculation
 as these stars contribute the most towards the total ionizing luminosity of an OB association.
 Since the least massive O stars with $\sim 20$ M$_{\odot}$ have a life time of $\sim 10$ Myr (see the discussion
 before eqn \ref{eq:mechlum}), we have used a lower mass cutoff for supernovae at $20$ M$_\odot$
 to calculate the value of $N_{O}$ for a given cluster mass. This is also consistent with the relation
 of mechanical luminosity with $N_{O}$ in eqn \ref{eq:mechlum}.
\begin{figure}
\centerline{
\epsfxsize=0.47\textwidth
\epsfbox{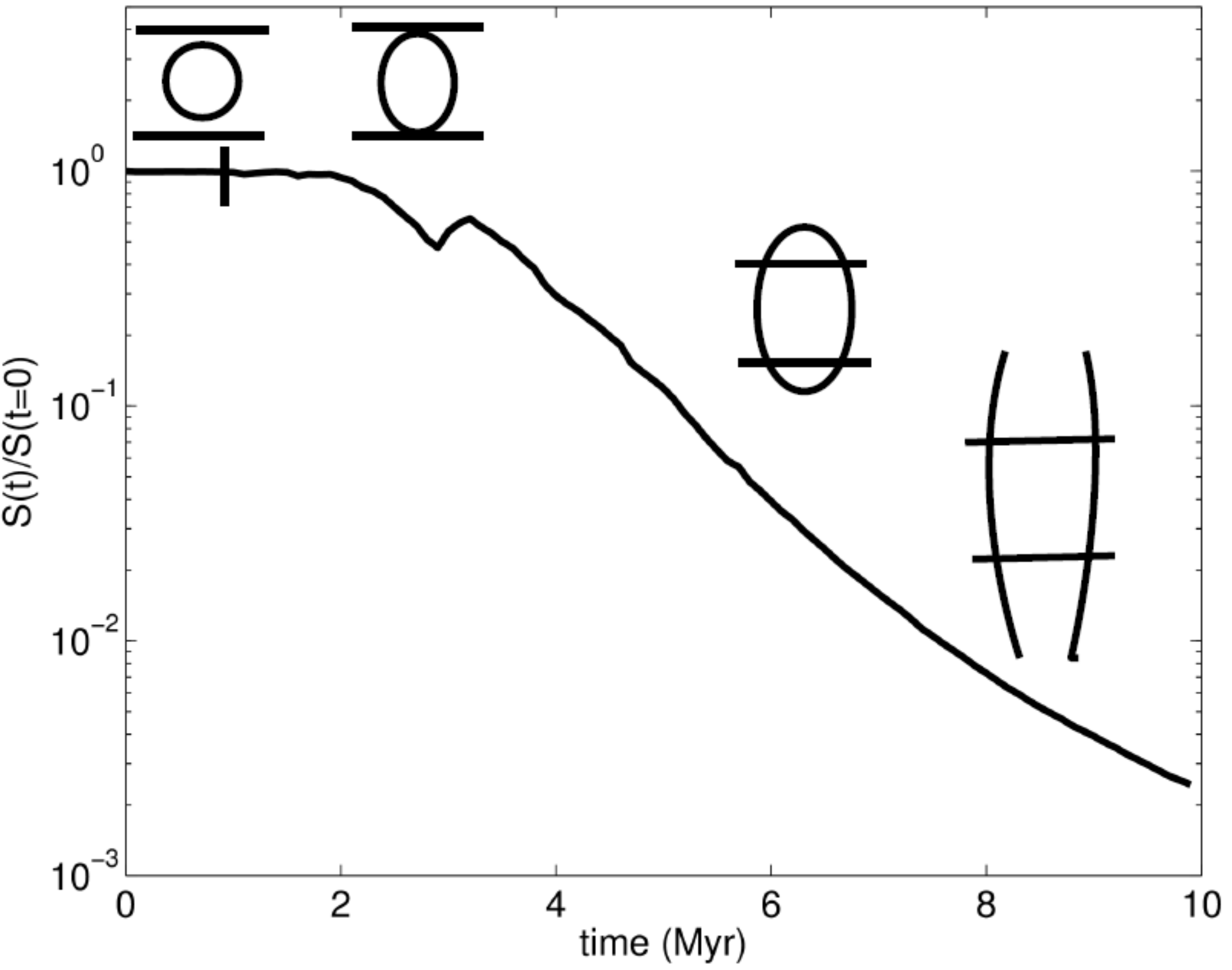}
}
\caption{
Normalized LyC photon luminosity as a function of time for a starburst calculated
using Starburst 99.
The dynamical time scale (of superbubble shells reaching the scale height)
for $n_0=0.5$ cm$^{-3}$, $z_0=300$ pc ranges between $0.4\hbox{--}4.2$ Myr for different $N_{O}$. For these values, we also
sketch the superbubble shells vis-a-vis the disk, beginning from the left with a small spherical shell, then with an elliptical shell slowly breaking out  
and finally ending with a shell whose top has been blown off by instabilities. The short vertical line at 0.4 Myr corresponds to the dynamical time ($t_d$) for $N_O=10^5$.
}
\label{fig:schem_diag}
\end{figure}
The ionizing photon luminosity is initially
constant when all the O-stars are present in the main sequence, and it decreases abruptly after 3 Myr as the sources of ionizing photons (O-stars)
start to die off. We plot a representative sketch of ionizing luminosity as a function
of time 
in Figure \ref{fig:schem_diag}. 
We also sketch the different evolution stages of the
superbubble. Initially the superbubble shell is completely buried inside the disk, and then it takes an elliptical shape. After that, the shell breaks out of
the disk, and in the final stages the superbubble ends up with a shell whose top has been blown off by thermal instabilities and RTI.

\subsection{Escape fraction}
We calculate the escape fraction along different lines of sight varying the angle from 0 to $\pi/2$ 
(where we measure $\theta$ from the perpendicular to the disk), using  
eqn \ref{eq:esc_frac_1_theta}
at a given time and for a particular 
$N_{O}$. Then 
we average it over $4\pi$ steradian to get the $\theta$-averaged escape fraction as a function of time and $N_{O}$:
\begin{equation}
 \langle f_{esc} \rangle_{\theta} (t,N_{O})= {1 \over 4\pi}\left[2\int_0^{\pi/2}f_{esc}(\theta) \sin(\theta)d\theta \int_0^{2\pi}d\phi\right] \,.
 \label{eq:theta_avg_esc_frac}
\end{equation}
The time-averaged escape fraction for a given $N_{O}$ is then,
\begin{equation}
\langle f_{esc} \rangle_{\theta,t} (N_{O} ) = { {\int_0^{t_{O}} \langle f_{esc} \rangle_\theta (t, N_{O}) S(t) dt} \over {\int_0^{t_{O}} S(t) dt} } \,,
 \label{eq:time_avg_theta_avg_esc_frac}
\end{equation}
where t$_{O}=10$ Myr.

\begin{figure}
\centerline{
\epsfxsize=0.56\textwidth
\epsfbox{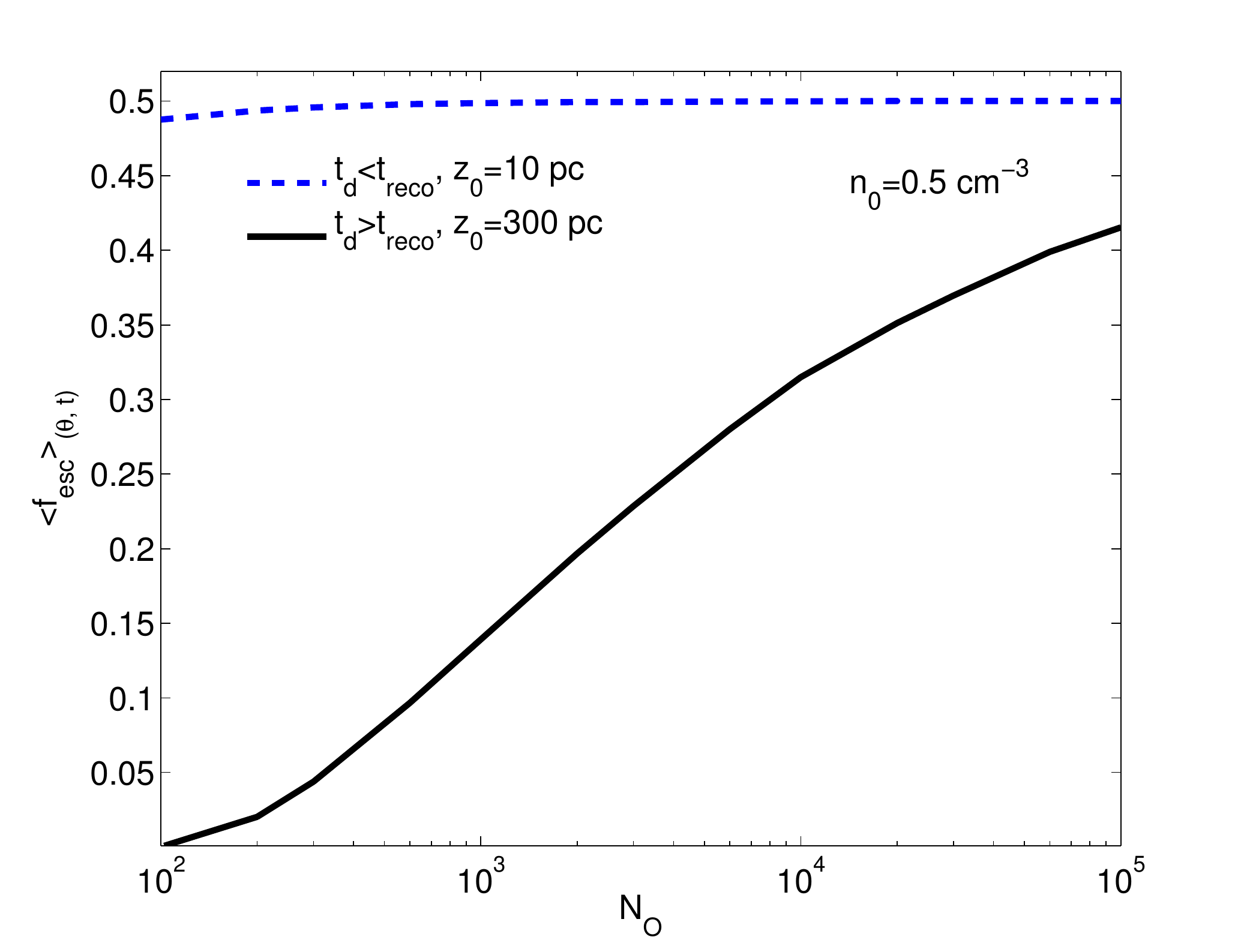}}
\caption{
 Time- and angle-averaged escape fraction as a function of the 
number of O stars for two scale heights,  including a smaller one for which the recombination time is longer than the dynamical time.
}
\label{fig:time_avg_theta_avg_fesc_nob}
\end{figure}

We plot this $\langle f_{esc} \rangle_{\theta,t}(N_{O})$
(eqn \ref{eq:time_avg_theta_avg_esc_frac}) as a function 
of $N_{O}$ for $n_H=0.5$ cm$^{-3}$ and for two values of $z_0=10$ and $300$ pc in Figure
\ref{fig:time_avg_theta_avg_fesc_nob}. For $z_0=10$ pc, the dynamical time $t_d < t_{\rm reco}$, the
recombination time scale, and the escape fraction is independent of $N_{O}$ (blue  dashed line),
whereas for $z_0=300$ pc, it increases with $N_{O}$ (black solid line). In the first case, the superbubble
reaches the scale height before substantial recombination takes place in the shell, and the escape fraction is dominated by
the dynamics of the superbubble than by the recombination rate. 
Our formalism is not valid in the first case, as explained in section 3.1. 

Next we convolve  $\langle f_{esc} \rangle_\theta(t,N_{O})$ with the luminosity function of OB associations, as given by \citet{mckee1997}, 
\begin{equation}
 { {dn} \over {dN_{O}} } {dN_{O}} \propto {  1 \over {N_{O}^2}} {dN_{O}} \,,
 \label{eq:lum_fn}
\end{equation}
where the LHS denotes the number of OB associations with 
the initial number of O stars in the range $N_{O}$ and $N_{O}+dN_{O}$.
The $N_{O}$-averaged escape fraction as a function of time is defined as
\be
 \langle f_{esc} \rangle_{\theta,N_{O}}(t) = \frac{\int_{N_{O_{1}}}^{N_{O_{2}}} \langle f_{esc} \rangle_\theta (t,N_{O}) S(t) {{dn}\over{dN_{O}}} dN_{O}}
{\int_{N_{O_{1}}}^{N_{O_{2}}} S(t) {{dn}\over{dN_{O}}} dN_{O} }.
\ee
For Salpeter IMF, the number of O stars $N_O=0.3 N_{\rm OB}$, the total number of OB stars. We use a lower limit $N_{{O}_1}=100$, corresponding to the smallest star clusters 
observed by \citet{zinnecker1993}, and an upper limit of $10^5$, for the largest clusters 
(\citealt{ho1997, martin2005, walcher2005}).

The angle and time dependence of the escape fraction from all the OB associations, averaging over only the luminosity 
function, is another important quantity. We define the luminosity-function-averaged escape fraction as:
\begin{equation}
   \langle f_{esc} \rangle_{N_{O}}(\theta, t)={ {\int_{N_{{O}_1}}^{N_{{O}_2}} f_{esc} (\theta, t, N_{O})S(t){{dn} \over {dN_{O}}} d N_{O}} \over {\int_{N_{{O}_1}}^{N_{{O}_2}}S(t){{dn} \over
   {dN_{O}}} dN_{O}}}\, .
   \label{eq:only_lum_avg_esc_frac}
\end{equation}
We use this definition in Figure \ref{fig:lum_f_theta}.
 
 Finally, the average 
 escape fraction takes the form,
  \begin{equation}
   \langle f_{esc} \rangle_{\theta,t,N_{O}}={ {\int_{N_{{O}_1}}^{N_{{O}_2}} \langle f_{esc} \rangle_{\theta,t} (N_{O}){{dn} \over {dN_{O}}} d N_{O}} \over {\int_{N_{{O}_1}}^{N_{{O}_2}}{{dn} \over
   {dN_{O}}}  dN_{O}} },
   \label{eq:final_esc_frac}
  \end{equation}
which is equivalent to $\int \langle f_{esc}\rangle_{\theta, N_{O}} (t) dt/t_{O}$.

\section{Results}
\label{sec:results}
In this section we present our results, beginning with the angle and time dependence 
of the escape fraction.  
Then show the the escaping LyC luminosity as a function of time.
Later we show the most interesting 
result of our study, namely, that the escape fraction increases slightly with 
a decreasing halo mass and a decreasing redshift.
Finally we discuss the effects of clumpiness 
as applied to high redshift galaxies.
\begin{figure}
\centerline{
\epsfxsize=0.56\textwidth
\epsfbox{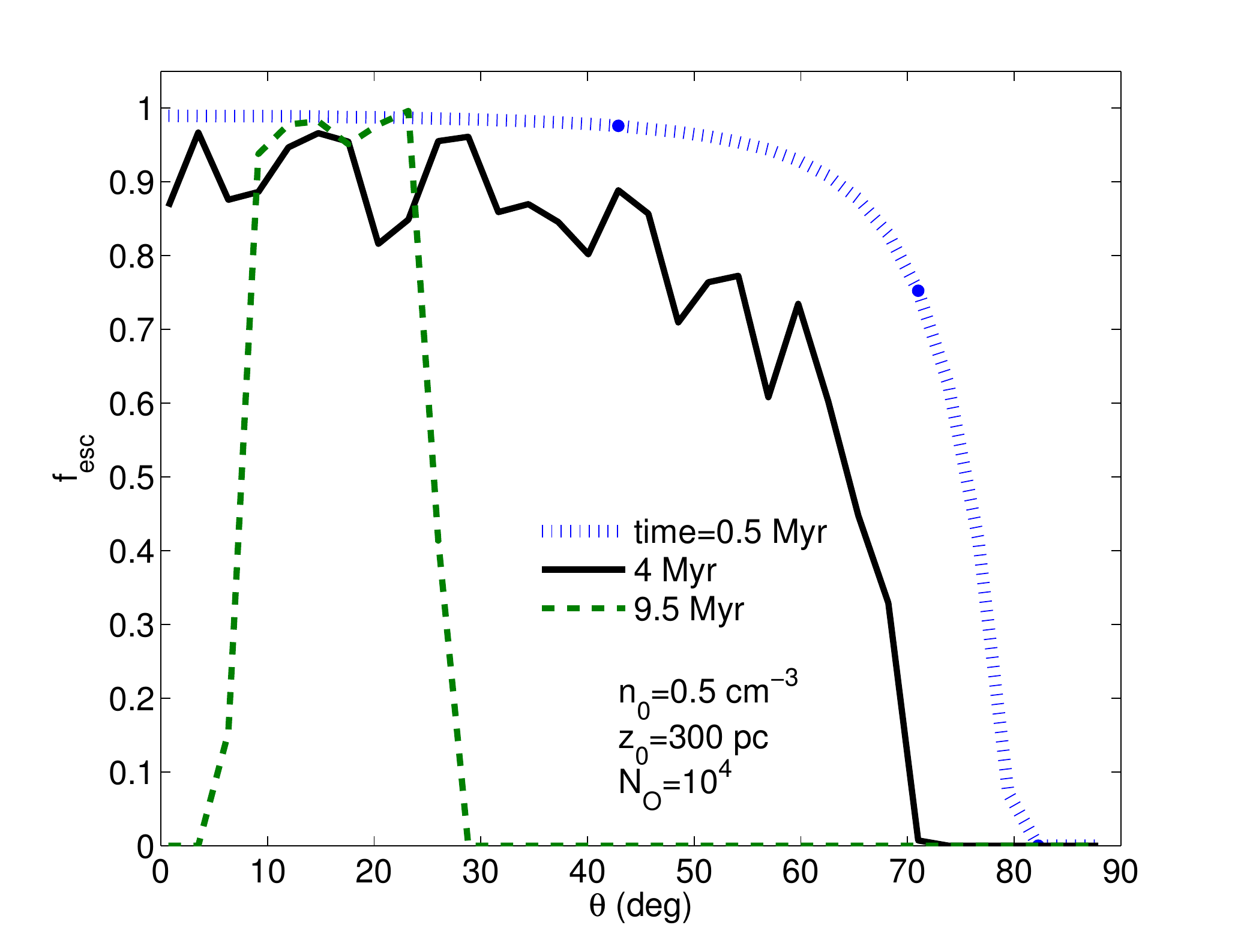}
}
\caption{
Escape fraction as a function of angle ($\theta$) at different times  (0.5, 4.0, 9.5 Myr) for $n_0=0.5$~cm$^{-3}$, $z_0=300$ pc, and $N_{O}=10^4$. The corresponding dynamical time $t_d\sim 1$ Myr. The blue dotted line represents the escape fraction
 at 0.5 Myr (at $t \ll t_d$, when the superbubble is deeply buried in the disk), the black solid line at 
 4 Myr ($t \approx 4 t_d$, when the superbubble shell begins to fragment, making the line zigzag), and the green dashed line at  9.5 Myr, 
 when the shell opens up completely at small angles. 
}
\label{fig:f_theta}
\end{figure}
\subsection{Angular dependence}
We first turn our attention to the angular dependence of the escape fraction for particular values
of the mid-plane density ($n_0$), scale height ($z_0$) and $N_{O}$.
In Figure \ref{fig:f_theta} we plot the escape fraction $f_{esc}(\theta)$ (eqn \ref{eq:esc_frac_1_theta}) 
as a function of $\theta$ 
at different epochs for $n_0=0.5$ cm$^{-3}$, $z_0=300$ pc, $N_{O}=10^4$.
The blue dotted line, black solid line, and green dashed line represent the angle dependence of the escape fraction at $0.5$, 
$4.0$, $9.5$ Myr respectively (times corresponding to the snapshots in Fig. \ref{fig:den_contour}). 
The escape fraction at all epochs slowly decreases with $\theta$ at small angles from the poles 
and then drops sharply at a cut-off angle.  
The reason for the decrease is that at larger angles (away from the poles) the line of sight encounters more of 
the disk/accumulated gas, and there is more recombination, making it difficult for the ionizing photons to escape. 
At time scales  $t \ll t_d$, the superbubble is mostly spherical and 
buried deep in the disk, and the escape fraction decreases smoothly with angle,
since all lines of sight encounter approximately the same path length through 
the disk, until $\theta \ge 1$ radian. At these epochs, $S$ is  large, and 
therefore the maximum value of the escape fraction (at small $\theta$) is also 
large. 

At $t  \sim 2\hbox{--}3  t_d$, after the superbubble breaks out from the 
disk, the shell begins to fragment due to radiative cooling and Rayleigh-Taylor
 instability (RTI; \citealt{roy2013}). These clumps  give rise to the zigzag nature 
of the angular dependence of the escape fraction. At small $\theta$, the 
absorption of ionizing photons mostly occurs in the shell. At angles larger 
than $\sim 1$ rad, the escape fraction decreases rapidly, when the path of 
the photon begins to encounter the swept-up disk material.  At later times the 
ionizing luminosity  decreases ($S[t]$; see Fig. \ref{fig:schem_diag}), and therefore the maximum value of the
escape fraction (at small $\theta$) also decreases. At a much later epoch, 
$t \gg t_d$, the shell opens up completely at small angles, giving a boost 
to the maximum value of escape fraction. However, the further decline of the 
ionizing luminosity also ensures a rapid decrease in the escape fraction with 
angle.  We also notice a large drop in the escape fraction at small angles (near the $\theta$-boundary in our numerical simulation) at 9.5 Myr. 
This is due to artificial accumulation of cold/dense gas near the poles, which is a feature of spherical
geometry used for the simulation. Since the solid angle covered by these small angles is negligible, our results are not affected by this spurious behavior at the poles.

\begin{figure}
\centerline{
\epsfxsize=0.50\textwidth
\epsfbox{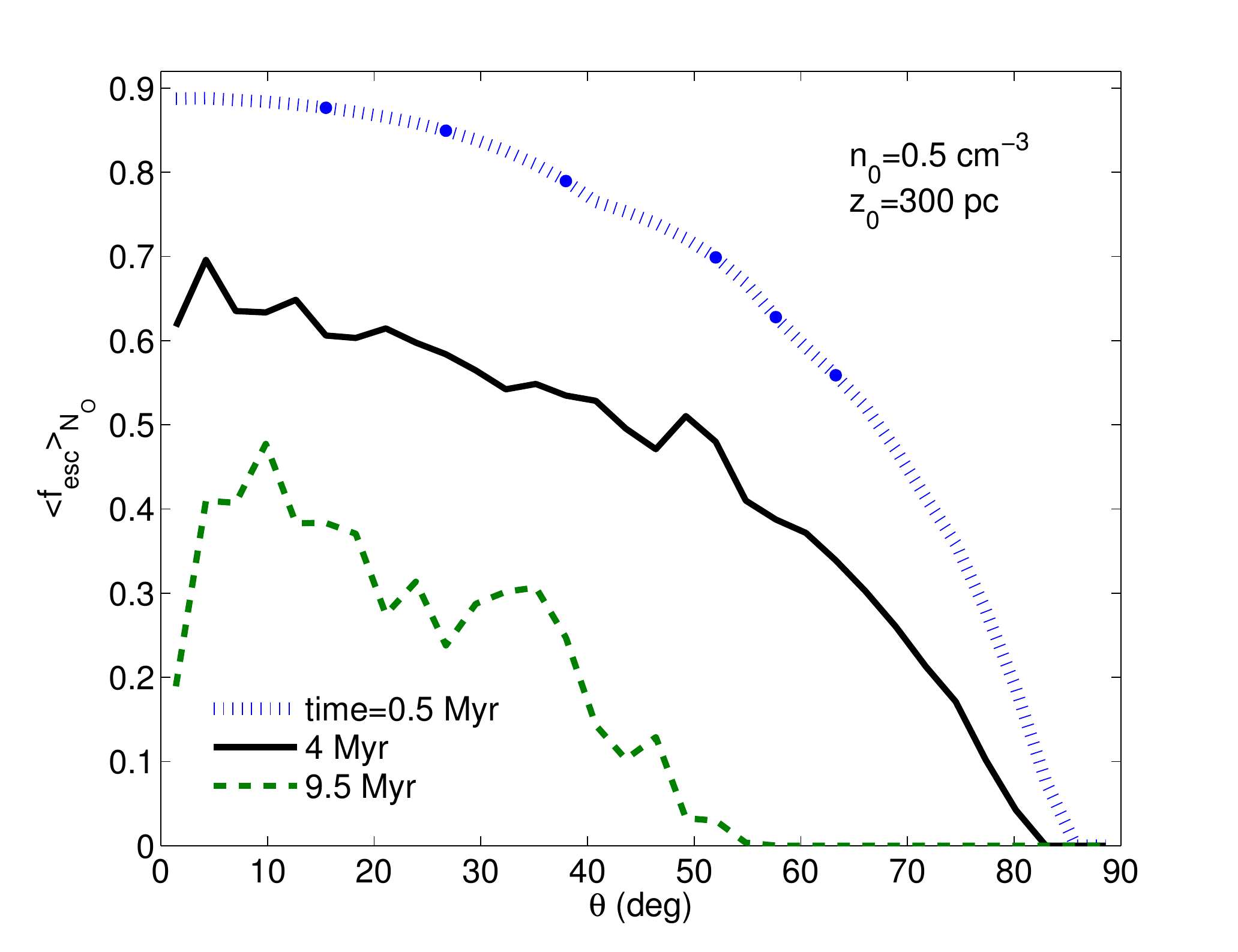}
}
\caption{
Luminosity-function-averaged escape fraction as a function of $\theta$ at different times (0.5, 
4.0, 9.5 Myr) for our fiducial disk ($n_0=0.5$ cm$^{-3}$, $z_0=300$ pc). The blue dotted, 
black solid and green dashed lines represent the cases at 0.5 Myr, 4 Myr, 9.5 Myr respectively.   }
\label{fig:lum_f_theta}
\end{figure}

The opening angle of the cone can be estimated from the Kompaneets 
approximation of an adiabatic superbubble \citet{kompaneets1960}. Consider cylindrical
geometry with ($R,z$) coordinates, and consider
the epoch when the superbubble has reached a height of $2 z_0$ in the $z$-direction.
The shell intersects the plane $z=z_0$ at some distance from the axis $r=0$. The perpendicular distance of the point of intersection of the shell with the plane $z=z_0$ is
 (see Eqs. 7 and 8 in \citealt{roy2013}),
 \be
 r=2z_0 \cos ^{-1} \Bigl [ {\sqrt{e} \over 2} [1- (1-1/e)^2 +1/e] \Bigr ] \,.
 \ee
 The angle that this point of intersection makes with the pole is
 \be
 \theta =\tan ^{-1} {r \over z_0}=\tan ^{-1} \Bigl [ 2 \cos ^{-1}  \Bigl ( {\sqrt{e} \over 2} 
 \Bigl ({3 \over e}-{1 \over e^2} \Bigr ) \Bigr ) \Bigr ] \sim 52 ^\circ \,,
 \label{komp_angle}
 \ee
  or roughly 1 rad, as expected from simple arguments mentioned above.

Next, we calculate the angular dependence of the escape fraction after convolving with the luminosity function of OB associations (using eqn \ref{eq:only_lum_avg_esc_frac}), for the same 
mid-plane density and scale height. Figure \ref{fig:lum_f_theta} shows the angular dependence at three different epochs 
(0.5, 4.0, 9.5 Myr, shown
with blue dotted, black solid, green dashed lines). The convolution with luminosity function ensures a greater contribution from associations 
with lower $N_{O}$, which leads to a marked decrease in the peak value of $f_{esc}$ with time, compared to that seen in Figure 
\ref{fig:f_theta}.

For the parameters used in  Figure \ref{fig:lum_f_theta}, the dynamical time 
$t_d\approx 2$ Myr for the dominant $N_{O}=600$ (see Appendix B for the estimate of dominant $N_{O}$). The blue dotted line shows the angular dependence at 
$t\ll t_d$, when the shell is approximately spherical and small. The solid 
black line shows the case when the superbubble has broken out of the disk and 
has started fragmenting due to RTI. The green dashed line shows the case at a 
much later epoch
($t\gg t_d$) 
with a rapid fall with angle, as explained above.

\begin{figure}
\centerline{
\epsfxsize=0.52\textwidth
\epsfbox{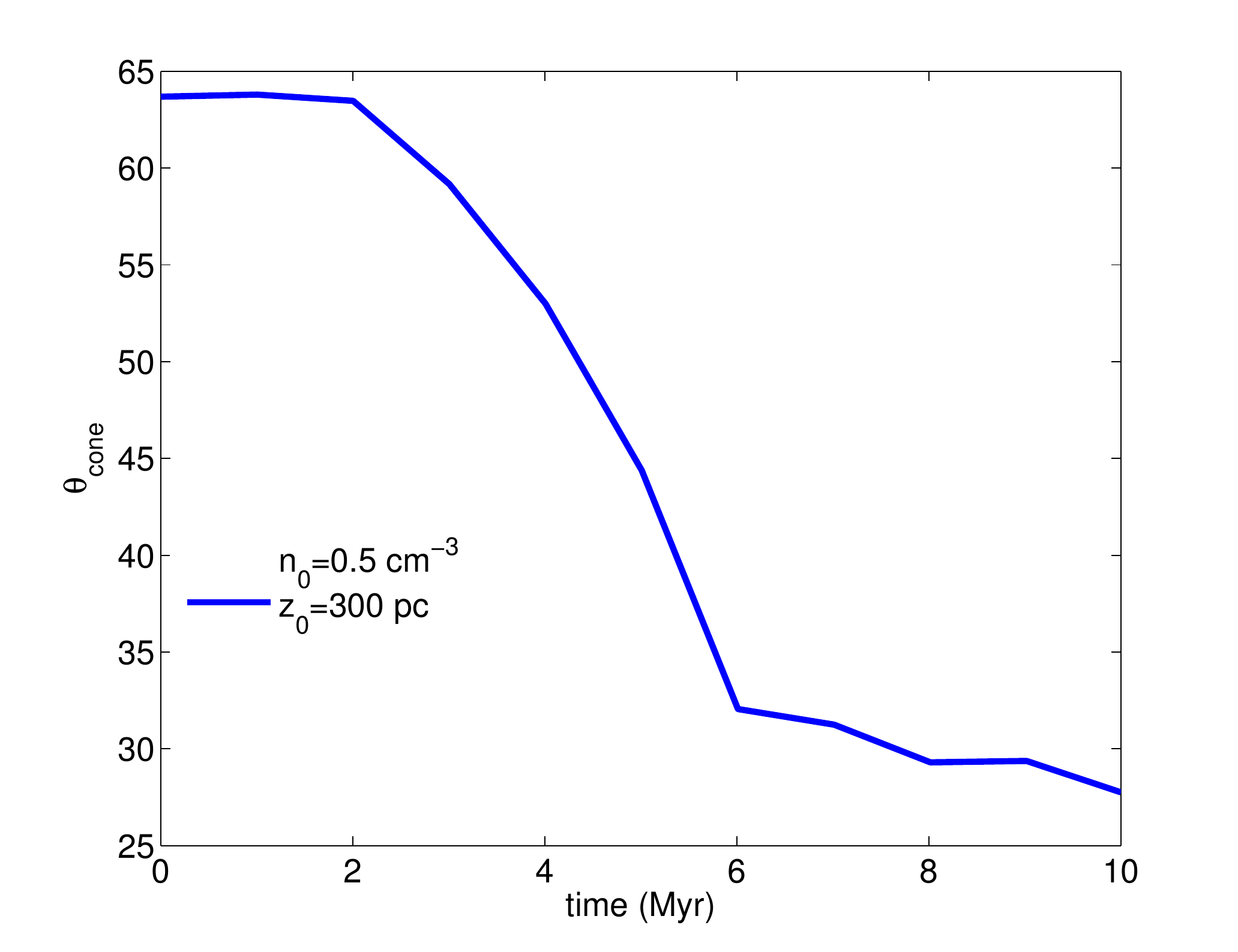}
}
\caption{
The time evolution of the ionization cone opening angle ($\theta_{\rm cone}$; where escape fraction falls by $[1-1/e]$ of its peak value) 
for our fiducial disk ($n_0=0.5$ cm$^{-3}$ and $z_0=300$ pc).}
\label{fig:theta_time}
\end{figure}
Defining the cone angle $\theta_{\rm cone}$ as the angle  at which the value of $f_{esc}$ drops to $(1-e^{-1})$ of
its peak value, we plot $\theta_{\rm cone}$ as a function of time in Figure \ref{fig:theta_time}.  
We find the cone angle to be $\sim 55\hbox{--}65^\circ$ 
(of order 1 rad) at $t\le t_d$, as expected from the above discussion in eqn \ref{komp_angle}. After superbubble breakout, $\theta_{\rm cone}$ declines rapidly, and reaches an asymptotic 
value $\sim 40^\circ$. This is because of opening of the top part of the shell due to instabilities. 
We note that the cone angle does not represent 
just the opening up of the superbubble shell as it evolves (as shown in Figure \ref{fig:den_contour}); it is also affected by the decreasing ionizing luminosity with time.
This is how we reconcile a decreasing $\theta_{\rm cone}$ in Figure \ref{fig:theta_time} with an increasingly wider superbubble seen in Figure \ref{fig:den_contour}
as time progresses.

We note that
our result is consistent with the results of \citet{fujita2003}, who found an opening angle of $30\hbox{--}40^\circ$ of the superbubble at 
$\sim 6$ Myr (after the shell fragments due to RTI) in their simulation of dwarf galaxies. 
Our result is also 
consistent with the recent observations of \citet{zastrow2013}, who found an average cone angle of 
$\approx 40^\circ$ in the case of six nearby dwarf starburst galaxies. In a survey of 6 Lyman Break Galaxies and 28 Lyman-$\alpha$ Emitters 
at $z\sim 3$, \citet{nestor2011} concluded that LyC photons escape over a fraction $0.1\hbox{--}0.2$ of the total solid angle, which implies a 
cone angle of $30\hbox{--}40^\circ$.

The fact that the opening angle is never larger than about 1 radian for a disk galaxy (eqn \ref{komp_angle}) also leads to 
an important conclusion. If we only consider the geometric effects on the escape of ionizing photons, and assume
that all ionizing photons inside the cone manage to escape, and those outside of it do not, then the maximum escape fraction from a disk galaxy is 
$\sim (1-\cos [1~{\rm radian}])=0.5$. As we have discussed in the Introduction, this inevitable limit appears to be borne out by the observations of $f_{\rm esc}$.

\subsection{Time dependence of average escape fraction}
The time-dependence of the escape fraction is governed by the competition between dynamical evolution of the superbubble which opens up an
ionization cone and the lifetime of O stars. If dynamical time is short, O stars are still around at the time of opening up of the superbubble and the 
escape fraction can be high.     
 
\begin{figure}
\centerline{
\epsfxsize=0.55\textwidth
\epsfbox{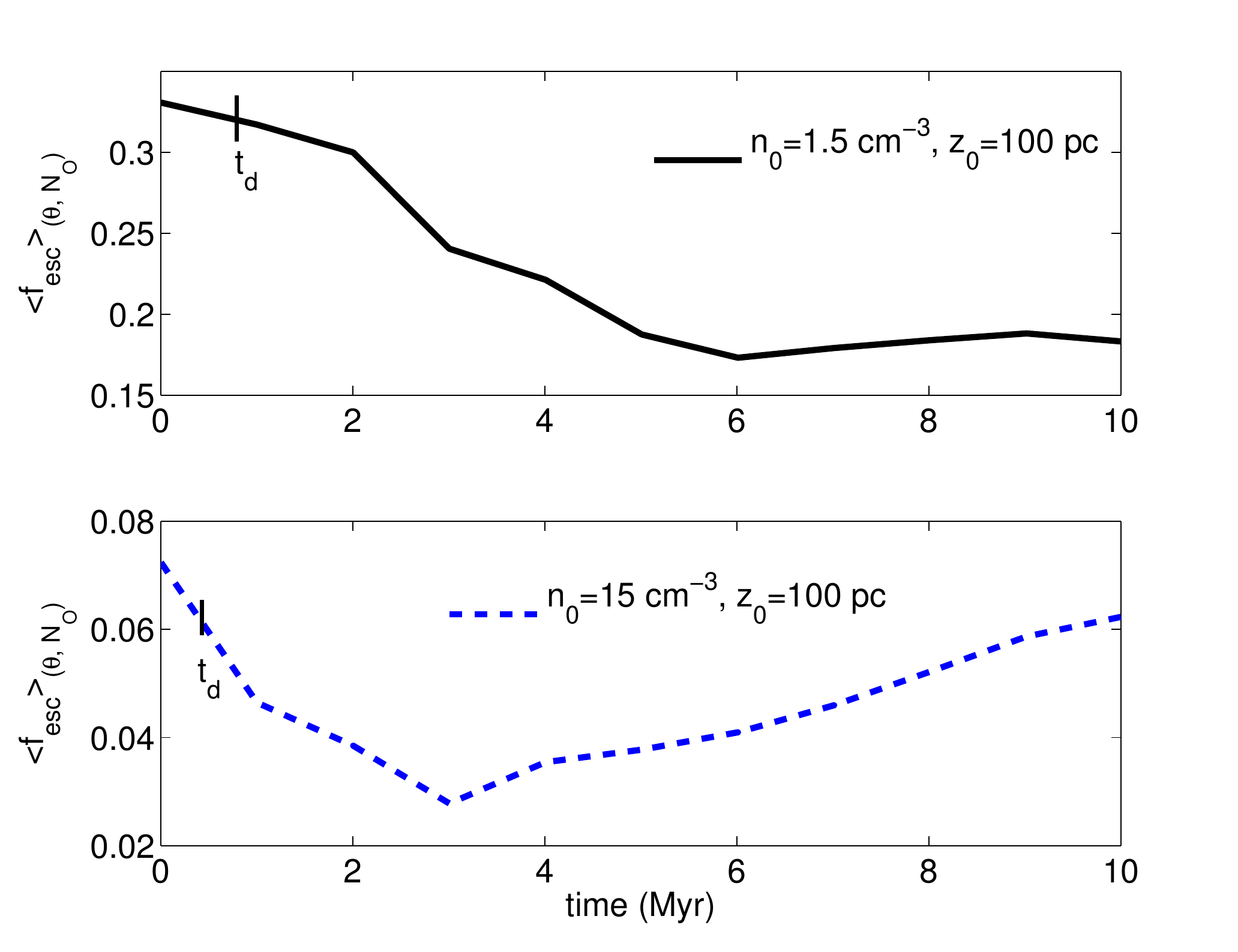}
}
\caption{
Luminosity-function-averaged $\theta$-averaged escape fraction as a function of time for two different
$n_0$ (1.5, 15 cm$^{-3}$) and a particular $z_0=100$ pc, to show the difference in their overall behaviour as a function time.
The dynamical time ($t_d$) for the two cases is marked with a short vertical line.}
\label{fig:fesc_time_high_low_n0}
\end{figure}

Figure  \ref{fig:fesc_time_high_low_n0} shows the luminosity function and $\theta$-averaged escape fraction as a function of time for 
$z_0=100$ pc and for two different mid-plane densities $n_0=1.5,15$ cm$^{-3}$.  
 Initially the superbubble is small
compared to the scale height, and the escape fraction depends on the magnitude of recombination in the dense shell 
of the superbubble (greater recombination reduces the escape fraction). As the superbubble grows
in size, it sweeps up more and more of the ISM gas, the thickness of the shell increases, and the escape fraction decreases. After the 
superbubble breaks out of the disk, and the top of the shell opens up due to Rayleigh-Taylor instability (RTI), the escape fraction increases 
if the O stars are still around.

The time scale of the onset of the RTI depends on $N_{O}$, or equivalently, the mechanical luminosity driving
the superbubble. For $n_0=1.5$ cm$^{-3}$, superbubbles with $N_{O}\approx 200$ dominate (see Appendix B for a discussion of dominant $N_{O}$)
the integration in eqn 
\ref{eq:final_esc_frac}, for which $t_d\sim 1$ Myr (marked in the figure). The onset of RTI takes place at $4\hbox{--}5$ times the dynamical 
time (\citealt{roy2013}), and therefore
the escape fraction keeps decreasing until $(4\hbox{--}5) \times t_d$. In the case of a denser medium, the
integration in eqn \ref{eq:final_esc_frac} is dominated by larger superbubbles, in this case with
$N_{O}\approx 3\times10^4$, for which $t_d\approx 0.4$ Myr. Since the dynamical time is shorter in this
case, the escape fraction reaches a minimum value at an earlier time, as shown in the bottom panel
of Figure \ref{fig:fesc_time_high_low_n0}.

  Since the life times of the most massive O stars lie in the range 
$2\hbox{--}4$ Myr, if the superbubble shell breaks much later than this,  the escape fraction does not increase much at later 
epochs (top panel of Figure \ref{fig:fesc_time_high_low_n0}). \citet{zastrow2013} found that the optimal time scale for the escape of LyC 
photons is $3\hbox{--5}$ Myr, before which the superbubbles have not broken out of the disk, and after which the LyC luminosity starts declining. 
This is consistent with our results.

\begin{figure}
\centerline{
\epsfxsize=0.55\textwidth
\epsfbox{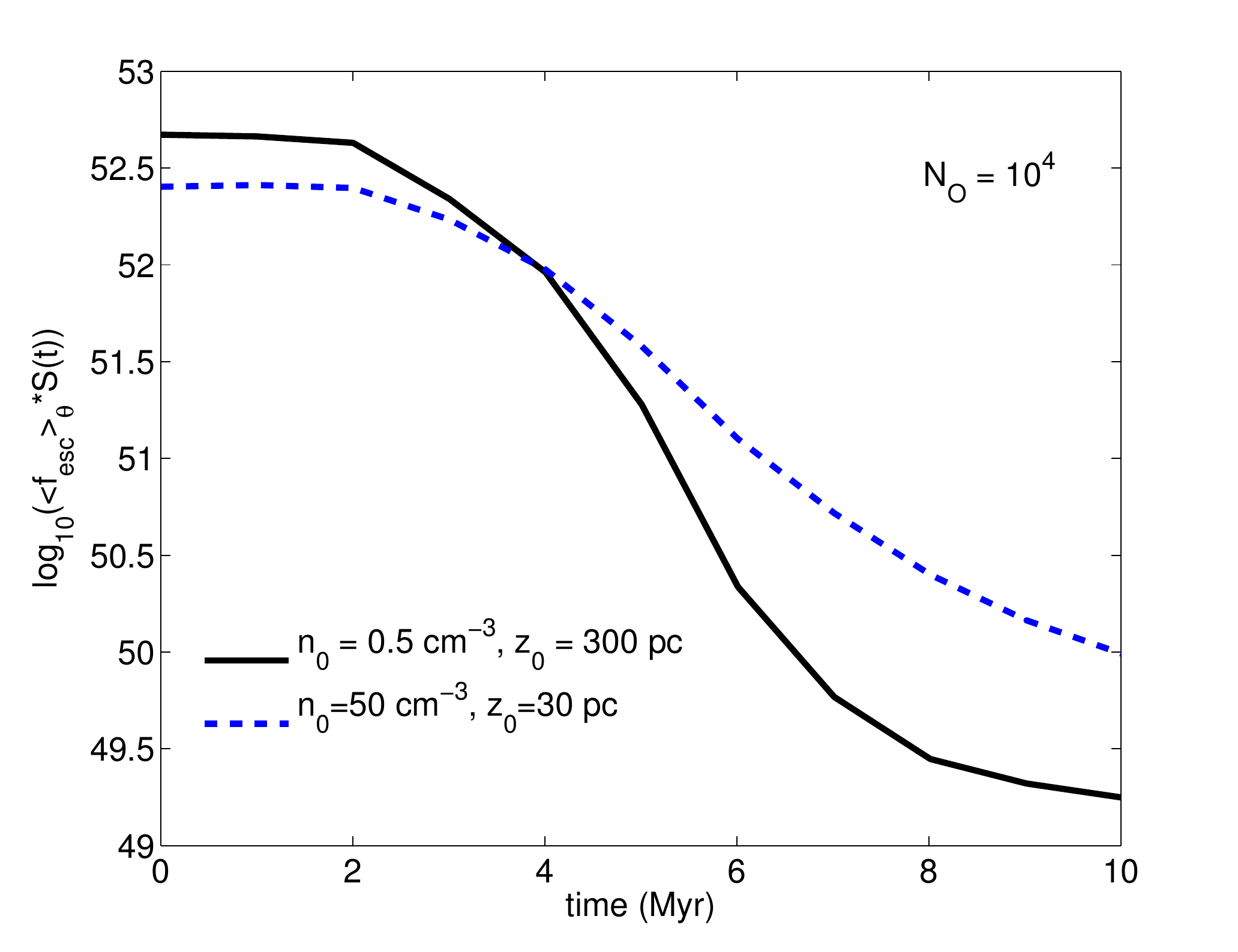}
}
\caption{
Time variation of the escaping ionizing photon luminosity from a star-cluster for two cases: $n_0=0.5$ cm$^{-3}$, $z_0=300$ pc, $N_{O}=10^4$;
and $n_0=50$ cm$^{-3}$, $z_0=30$ pc, $N_{O}=10^4$.}
\label{fig:flux_time}
\end{figure}

It is also important to study the final emergent ionizing luminosity ($f_{\rm esc} \times S(t)$), since this is what is 
observed. Figure \ref{fig:flux_time} shows the evolution of the ionizing luminosity for $N_{O}=10^4$, for the fiducial
disk parameters ($n_0=0.5$ cm$^{-3}$, $z_0=300$ pc), and for a higher density ($n_0=50$ cm$^{-3}$) and a smaller scale-height (30 pc) case. 
Disk galaxies produce most of their ionizing photons in the initial $\le 3$ Myr,
after which the total ionizing output decreases rapidly. Although $f_{\rm esc}$ increases at later epochs ($\sim 10 $ Myr)
due to opening up of the shell,
the observed ionizing luminosity is typically small because the decrease in photon luminosity (Fig. \ref{fig:schem_diag}) is much more than the 
increase in the escape fraction (Fig. \ref{fig:fesc_time_high_low_n0}). This is also borne out in the observations of SMC and LMC by \citet{pellegrini2012} who found
escape fractions from individual cluster regions to be $\sim 0.4$, but which had low ionizing luminosities. A comparison of the emerging photon flux
for the fiducial and the higher density cases shows that the emerging flux can be larger at later times for the higher density disk, even though 
the time-averaged escape fraction is lower (see Fig. \ref{fig:fesc_n0_z0}). This can be understood in terms of the propagation of the ionization 
front and the opening up of the low density cone due to supernovae. The escape fraction is initially smaller for the 
high density case as  $n_0^2 z_0^3/S$ is larger, but the superbubble breaks out faster (see eqn \ref{tdyn}) for this case because of a smaller scale height.

 \subsection{Average escape fraction}
\begin{figure}
\centerline{
\epsfxsize=0.55\textwidth
\epsfbox{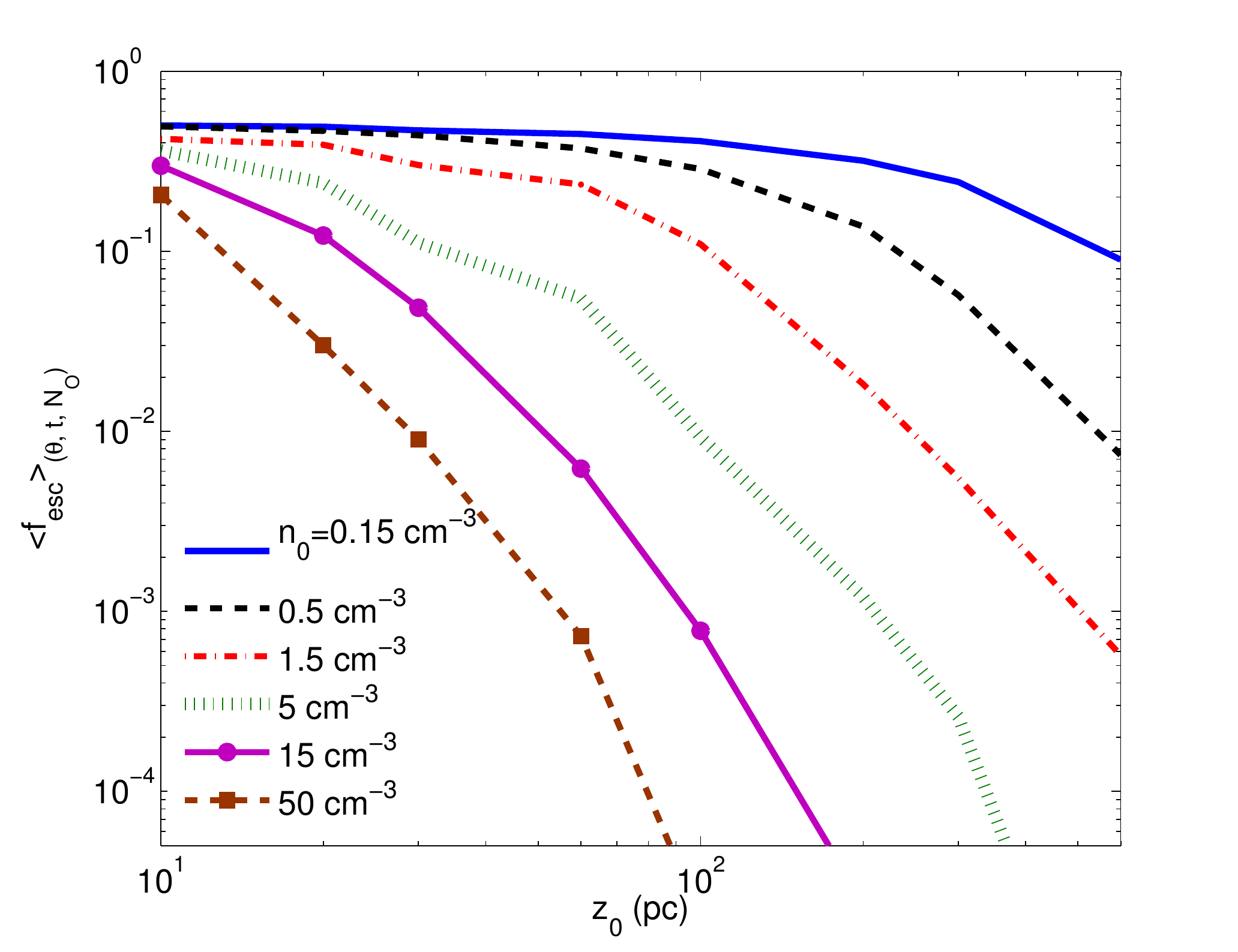}
}
\caption{
Time-averaged, luminosity-function-averaged  and $\theta$-averaged escape fraction as a function of $z_0$ for 
different $n_0$. Notice the sharp fall in the escape fraction with an increase in $z_0$.}
\label{fig:fesc_n0_z0}
\end{figure}

The most important quantity for reionization and UV background is the average escape fraction, averaged over all angles, times and the OB-association
number distribution. The angular and time dependence of the escape fraction is more relevant to understand the escape of UV photons in a particular system.

Figure \ref{fig:fesc_n0_z0} shows the average escape fraction  as a function of $z_0$ for different $n_0$. The curves show that with an increase in gas density,
the decrease of the escape fraction with scale height becomes sharper. For disks with small density and scale
height, the average escape fraction can be as larger as $\sim 0.5$, while for $n_0=0.5$ 
cm$^{-3}$ and $z_0=300\textendash 500$ pc, similar to the Milky Way disk, the escape fraction is $\sim 0.05$. 
However, the cases with small densities and small scale heights need to be considered with caution, as explained below.
 
\begin{figure}
\centerline{
\epsfxsize=0.55\textwidth
\epsfbox{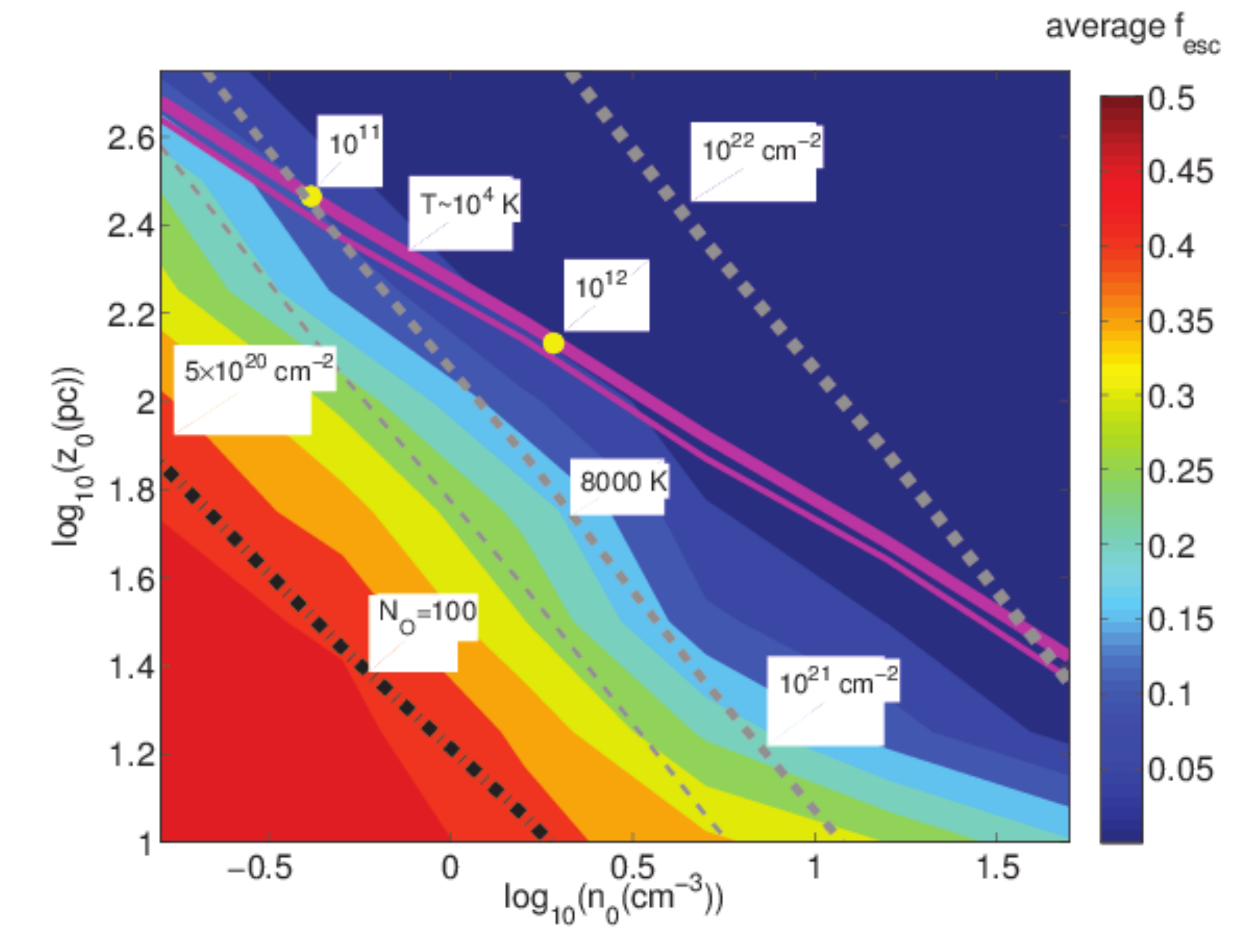}
}
\caption{
Contour plot of time-averaged luminosity-function-averaged and $\theta$-averaged escape fraction as a function of $n_0$ and $z_0$.
The regions below the black dashed-dotted  line is for $t_d<t_{\rm reco}$ for 
 $N_{O}=100$. The magenta solid thick and thin  lines represent the $n_0$, $z_0$
values for two different ISM temperatures $10^4$K, 8000 K respectively of the warm neutral medium (WNM). The yellow circular scatterers
represent the $n_0\textendash z_0$ values corresponding to the density and scale height calculated from
  Wood \& Loeb (2000) for the halo masses of 
 $M_h\sim 10^{12} M_{\odot}$ (the lower circle) and $M_h\sim 10^{11} M_{\odot}$ (the upper circle) respectively at present redshift ($z=0$).
 The grey dashed thick, thinner and thinnest lines represent constant HI-column density of $N_{\rm HI} \sim 10^{22}$ 
cm$^{-2}$, $10^{21}$ cm$^{-2}$ and $5 \times 10^{20}$ respectively. 
}
\label{fig:contour_plot}
\end{figure}

Figure \ref{fig:contour_plot} is a contour plot of the average escape fraction as a function of logarithmic $n_0$ 
and $z_0$. The colour bar represents the value of the escape fraction. 
As is evident in Figure \ref{fig:fesc_n0_z0}, the regions in the parameter space of $n_0\textendash z_0$ with small density and small scale 
height show larger values of the escape fraction than at high density and large scale height.

A note on the dominant $N_{O}$ that contributes the most to the value of the escape fraction
is in order here. For small $n_0$ and $z_0$, near the bottom-left corner of Figure \ref{fig:contour_plot},
the dominant value of $N_{O}$ is $\approx 100$, and superbubbles with larger $N_{O}$
start dominating as one goes towards the top-right corner (large $n_0$ and $z_0$). For $n_0=5$ cm$^{-3}$ and $z_0=100$ pc, the 
dominant $N_{O}=4000$ (see Appendix B for a discussion of the dominant $N_{O}$ as a function of the disk parameters $n_0$ and $z_0$).
The dominant $N_{O}$ corresponds to the value for which the integrand in the numerator of eqn \ref{eq:final_esc_frac} peaks.

Recall the discussion on the validity of the assumption of ionization balance in \S 3.1. Our formalism is not valid if 
$t_{\rm reco} > t_{\rm d}$, and this inequality depends on $N_{O}$ for a given combination of disk density and scale height.
We have found that this problem arises for small values of $N_{O}$,  especially 
at the bottom-left corner of Figure \ref{fig:contour_plot} for small $n_0$ and $z_0$. In this region, the dominant 
$N_{O}=100$ and $t_{\rm reco} > t_{\rm d}$. We show with a black dashed-dotted line the locus of 
points with $t_{\rm reco} =t_{\rm d}$ 
for $N_{O}=100$. The results for escape fraction for the region on the left of this line are not strictly valid.  
On the right hand side of this line, the dominant $N_{O}$ is such that 
$t_{\rm reco} < t_{\rm d}$, and our results are valid. We note that  photoionization/recombination equilibrium holds  
for most $n_0$ and $z_0$ considered here. Also, equilibrium disks with these combinations
of density and scale height correspond to very low ISM temperatures (in the range of $150\hbox{--}4000$ K, see eqn \ref{eq:z_distribution}), 
and should not be considered realistic.  

The grey dashed thick, thinner and thinnest lines in Figure \ref{fig:contour_plot} represent constant column densities of $10^{22}$,  $10^{21}$ and $5\times 10^{20}$ cm$^{-2}$, respectively, for a vertical 
line-of-sight.
The column density is given by,  
\begin{eqnarray}
N_{H} &=&n_0\int_{-\infty}^{\infty} {\rm sech}^{2}\bigl({z \over {\sqrt{2}z_0}}\bigr)dz = 2\sqrt{2} n_0 z_0. \, \nonumber\\
 \label{eq:column_den}
\end{eqnarray}
We note that at redshift $z\sim 0$, disks with $N_{\rm HI}\sim 10^{21}$ cm$^{-2}$ dominate 
the mass
 density of HI (\citealt{zwaan2005}), and possibly also at high redshifts. 
  \citet{prochaska2005} however suggests that disks with $N_{\rm HI}\sim 10^{20.3}$ and 
 $10^{21.3}$ contribute more or less similarly in
 the overall mass-density of HI. 
It has been pointed out by several authors (\citealt{cen2012, schaye2001, zwaan2006, hirashita2005}) that at $z\sim 0$ the systems above 
 $N_{\rm HI} \sim 10^{22}$ are very difficult to find due to HI-H2 coversion.  
In general, the escape fraction is lower for low column density systems, as expected.

The magenta solid thick and thin  lines in Figure \ref{fig:contour_plot} correspond to  the ISM temperatures 
of $10^4$ K and 8000 K respectively. \citet{wolfire2003} %
considered the thermal and ionization balance in the Milky Way ISM,
and inferred a range in the disk temperature in which two phases can coexist to be
$T\sim 7000\hbox{--}8500$ K.
We show the lines for disks with ISM temperatures of   
8000 K and $10^4$ K according to eqn \ref{eq:HSE}
The escape fraction
from disks at higher temperature ($T\sim 10^4$ K) is lower than that at lower 
temperature ($8000$ K). As we see from eqn  \ref{eq:HSE},  higher temperature corresponds 
to a larger scale height (for a given density) due to the puffing up of the disk.
 In that case, it becomes difficult for ionizing photons to escape the disk, and thus explains the 
 behaviour of $f_{\rm esc}$ with the disk temperature.

It is important to note that this density-height relation is roughly independent of the disk (galaxy) mass and redshift.
 The WNM disk column increases with the galaxy mass and redshift, but the relation between the mid-plane density and the
scale height satisfies eqn \ref{eq:HSE}, as long as the disk temperature remains the same and the disk is in hydrostatic equilibrium. 

Finally we come to the main result of our paper. {\it The contours of equal values of escape fraction
roughly obey the relation $n_0^2 \propto z_0^{-3}$.} In other words, disks in which the mid-plane density and scale height are related
such that $n_0^2 z_0^3$ is constant, would have similar escape fractions. Note that in the case of a constant ionizing luminosity, the
Str\"omgren sphere has a radius $R_s \propto n_0^{-2/3}$. This means that, the disks in which the Str\"omgren radius for a constant
ionizing luminosity is a fixed ratio of the scale height, would have similar values of escape fraction. In hindsight, one could argue that 
this is expected from a simple theoretical argument, because the escape fraction must depend on the amount of ionizing photons
absorbed by the disk gas, and so the ratio of the Str\"omgren radius to the scale height must be a relevant parameter. But without the aid of detailed calculation such as presented here, one could not have drawn such a conclusion with confidence, since there are a large
number of competing factors (such as opening up of low density channels) at play.

Note that lines of isothermal disks ($n_0 \propto z_0^{-2}$; eqn \ref{eq:HSE}) and those of constant column density ($n_0 \propto z_0^{-1}$; 
eqn \ref{eq:column_den}) straddle
the iso-$f_{\rm esc}$ contours from two sides in Figure \ref{fig:contour_plot}. In other words, $f_{\rm esc}$ of disks with constant WNM temperatures or disks with a given column densities would differ slightly. Consider disks with similar WNM temperatures that lie on the solid thick line in Figure \ref{fig:contour_plot}.
Disks of galaxies with different masses would be separated on this line, given by eqn \ref{eq:wl} (explained later in \S 4.5). We show two such points corresponding to halo masses $10^{11}$ and $10^{12}$ M$_{\odot}$ at the present epoch. The escape fraction for the disk of the less
massive galaxy is slightly larger than that of the more massive one (by a factor $\sim 1.4$).

Roughly one can fit  the values of the escape fraction to the mid-plane disk density and the scale height as,
\be
f_{\rm esc} \sim 0.1 \Bigl ( {n_0 \over 1 \, {\rm cm}^{-3}} \Bigr )^{-2/2.2} \Bigl ( { z_0 \over 135 \, {\rm pc} } \Bigr )^{-3/2.2} \,.
\ee
This fit is reasonably good for values of $n_0$ and $z_0$ for which
$ 0.133 \le \Bigl ( {n_0 \over 1 \, {\rm cm}^{-3}} \Bigr )^{2} \Bigl ( { z_0 \over 135 \, {\rm pc} } \Bigr )^{3} \le 7.5$ and $n_0\le 15$ cm$^{-3}$. Since gas
density and scale height depend on the WNM temperature (eqn \ref{eq:HSE}) and the galactic mass
(eqn \ref{eq:wl}), we can use this fit to  determine the dependence of $f_{\rm esc}$ on galactic mass and the WNM temperature.
Equation \ref{eq:wl} states that $n_0 \propto M^2/(c_s^2 R_d^4)$. Since $R_d$, the scale length, scales as the virial radius, which
scales as $M^{1/3}$, we have $n_0 \propto M^{2/3} c_s^{-2}$. Combining with the above fit and the relation $n_0 z_0^2 \propto c_s^2$, we finally have
\be
f_{\rm esc} \propto M^{-0.15} c_s ^{-0.9} \,.
\ee
Therefore, decreasing the mass of a galaxy by a factor of 10 increases the escape fraction by a factor $\sim 1.4$ (as also seen from the
two marked points in Figure \ref{fig:contour_plot}).

Present day disks occupy the upper-left corner of the parameter space in Figure \ref{fig:contour_plot}, shown by the
marked points for two halos. Disk galaxies therefore have an escape fraction of $f_{\rm esc} \sim 0.05\hbox{--}0.15$, with 
a weak variation with galactic mass.

This result admittedly pertains to discs that are not clumpy and we have not evolved the discs in a cosmological setting. However, our 
result does reveal an interesting connection between disc
parameters and the escape fraction. This interesting result could be obtained only after scanning a large parameter 
space of $n_0\textendash z_0$ that we have done here for the first time. In order to 
scan the parameter space, we used 48 different combinations of disc density and height, and each combination
had 13 runs with different $N_{O}$ for the averaging over OB associations. Our aim was to explore the effect of disc parameters in a 
simple case before introducing other effects such as clumpiness or cosmological and
galactic evolutionary effects. 

Another word of caution is in order here. Although we have drawn the points in Figure \ref{fig:contour_plot} for two halo masses, 
guided by the eqn \ref{eq:wl}, they
may not represent real disk galaxies with those halo masses. For one reason, the column densities in the prescription of \citet{mo1998}
are an overestimate. For example, the disk of our Milky Way (corresponding to a halo mass of $\sim 10^{12}$ M$\odot$) has a WNM column density of $\sim 2 \times (1.6 \times 10^{20})$ cm$^{-2}$ at high latitudes (implying a face-on geometry; Table 1 in \citealt{kanekar2011}),
almost an order of magnitude smaller than that predicted by \citet{wood2000} and two orders of magnitude smaller than
in the prescription of \citet{mo1998}. Therefore, the marked points in the figure should be used with caution, and we use
them here to demonstrate that low mass disks would in general occupy the upper-left corner of the $n_0-z_0$ space.

\subsection{Effect of clumping in the shell}
Clumpiness in the ISM can strongly affect the escape routes of ionizing photons in more than one way. Ionising photons escape easily
through the low density channels in the ISM. On the other hand, recombination rate is high in denser regions but the escape fraction also
depends on the photon luminosity.
Therefore it is not entirely clear how clumpiness should affect the 
angle-averaged, time-averaged and OB association luminosity-function-averaged escape fraction. 
Although there may not be a one-to-one 
correspondence between clumpiness (for a given definition) and the escape fraction, it is expected that $f_{\rm esc}$ increases with an 
increasing clumpiness (a higher clumpiness corresponds to the availability of low density escape paths for photons). 
\citet{wood2000} considered distributed sources in a disc and found that increasing the volume filling factor 
of clumps decreased the escape fraction. They found that the escape fraction is larger than $1\%$ only if the filling factor 
is $\le 0.2$, or if the interclump (empty) regions filled up more than $80\%$ of the ISM volume.

\begin{figure}
\centerline{
\epsfxsize=0.55\textwidth
\epsfbox{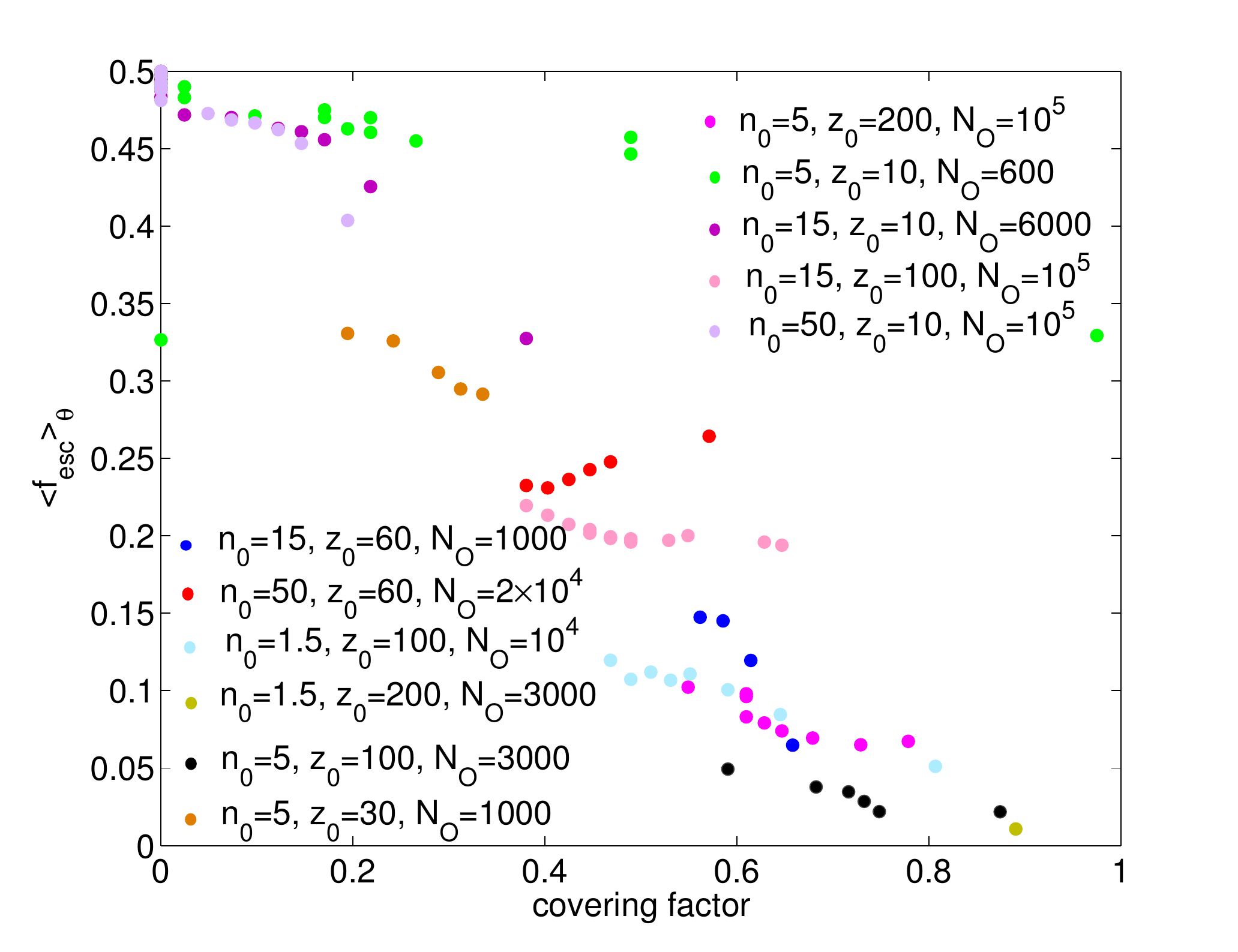}
}
\caption{
Escape fraction for a few cases (different $n_0, z_0, N_{O}$) are shown against the corresponding covering fraction (see text for details). All the 
densities are in cm$^{-3}$ and all the scale heights are in pc.
Different points in the plot correspond to the values of $\theta$-averaged $f_{\rm esc}$ and covering factor at different epochs of the superbubble 
evolution.
}
\label{fig:fesc_cover}
\end{figure}

Thermal and Rayleigh-Taylor instabilities give rise to clumps and channels in superbubble shell. These channels help the ionizing photons to 
escape and can 
increase the value of the escape fraction. Although we have not taken into account the
clumpy nature of the ISM, the clumps in the superbubble shell 
 mimic the clumpiness in the ISM. 
In this section we discuss how a clumpy ISM can affect the escape fraction, by studying the effect of the clumps in the shell on the
escape fraction. We show that an increased clumpiness in the medium (in the shell, and therefore, by extrapolation, in the ISM) decreases
the escape fraction.

Instead of the volume filling factors of clumps, which is difficult to infer in projected images of galaxies, 
we study the effect of surface covering factor on the escape fraction in our simulations. 
We define the covering factor as the fraction of the 
total surface area around the central source along which the line-of-sight column density is $\ge 10$ \% of  the initial value. We calculate 
the covering factor from the epoch when the escape fraction increases after reaching a minimum, when the shells are clumpy enough 
so that one can distinguish between clumps and channels in the shell. The column density along polar regions deceases as the 
superbubble evolves with time, since most of the volume is filled with the low density gas. 
 We have also 
checked the density contours at different snapshots visually, and found that our definition of the covering factor matches well the 
surface covering fraction estimated by considering dense clumps along different lines of sight.

We show a plot of the escape fraction in a few cases (gas density, scale height and $N_{O}$) as a function of the covering fraction 
in Figure \ref{fig:fesc_cover}. One can notice that there is a rough trend that as the covering factor decreases, the value of the escape fraction 
increases. It is clear that the escape of ionizing photons is facilitated 
by the opening up of pathways, which corresponds to a small covering fraction. For the escape fraction to be larger than $10\%$, the covering 
fraction needs to be less than $\sim 70\%$.

While we have used a particular definition of the covering factor (or clumpiness), in general we expect the escape fraction to rise with a decreasing
covering factor. An exception is when the UV sources are embedded inside the high density massive clumps (which is not the case, at least 
in our simulations).

\subsection{Variation with redshift}
\begin{figure}
\centerline{
\epsfxsize=0.58\textwidth
\epsfbox{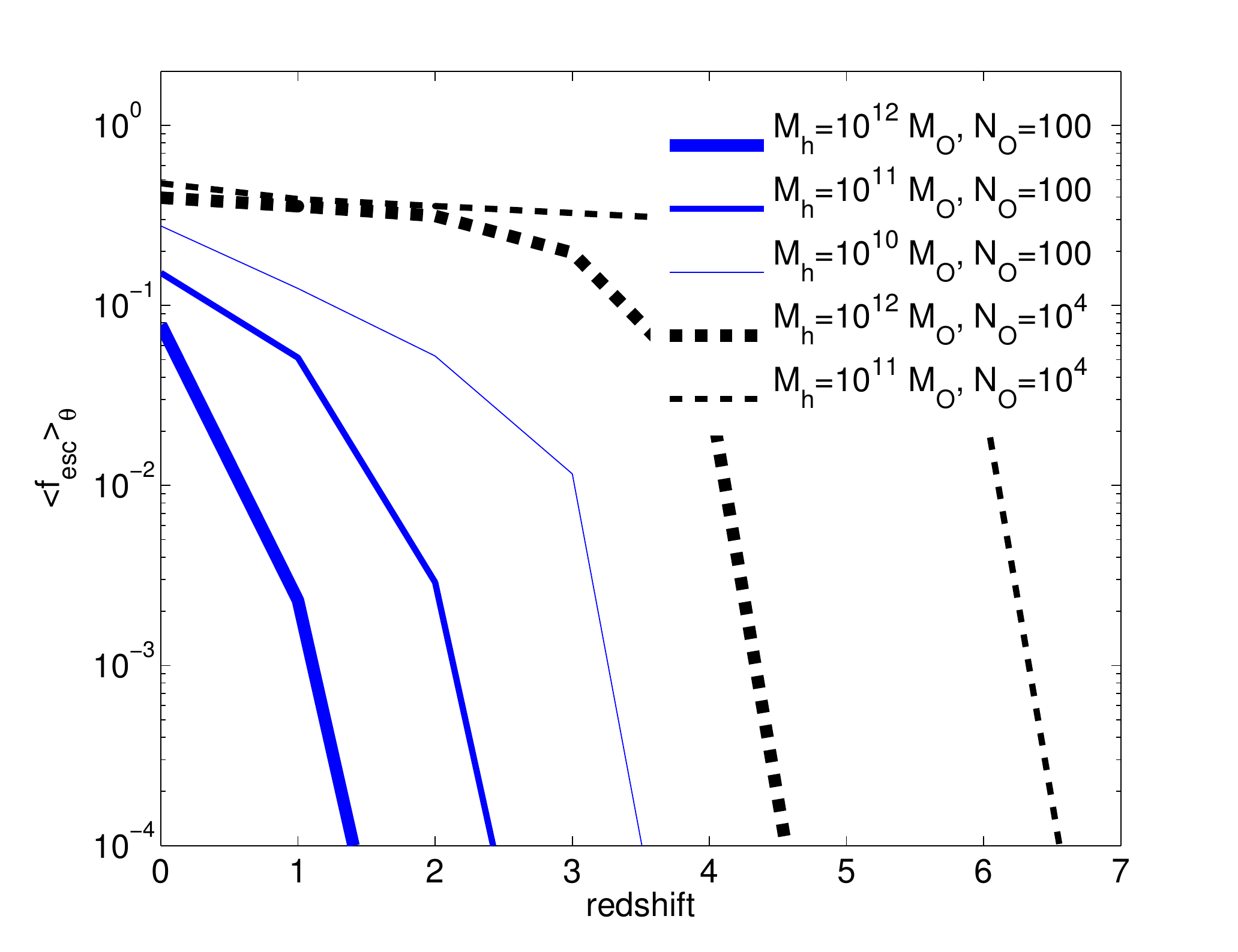}
}
\caption{
Escape fraction for the initial disk (without superbubbles) as a function of redshift, for a few halo masses and
$N_{O}=100, 10^4$. The purpose of this figure is to compare with the results of \citet{wood2000}.
}
\label{fig:fesc_z}
\end{figure}
One can use the
prescription of \citet{mo1998} to estimate the disk gas density for different galaxies at various
redshifts (see also \citet{wood2000}). Assume that the disk mass $M_d$ is a fraction $m_d$ of the halo
mass. Then, assuming a stratified disk with a (vertical) density profile given by eqn \ref{eq:z_distribution}, we have (eqn 8 
of \citet{wood2000})
\begin{equation}
 n_0={{GM_d^2} \over {128 \pi \mu m_p c_s^2 R^4}}\,,
 \label{eq:wl}
\end{equation}
where $ R$ is the scale radius of the mid-plane disk density (this is half of the scale radius corresponding to the 
surface density, which is used in \citet{mo1998}; this factor of two can make a factor of 16 difference in $n_0$ if one is not careful). 
 The redshift factor enters into 
this relation through the scale radius $R$, which is related to the virial radius $r_{vir}$, which depends on the redshift
(see eqn  5 of \citealt{wood2000}).  
One therefore obtains $n_0$ (and therefore $z_0$) for the disk of given halo mass at a certain redshift.
At high redshifts, disks with the same WNM temperature would slide down the solid lines in Figure \ref{fig:contour_plot} towards 
higher density and smaller scale height. Therefore in principle, one predicts smaller $f_{\rm esc}$ for disks with a given mass and temperature
at higher redshifts. However, this weak trend is likely to be mitigated by the stronger effect of clumpiness on the scape fraction, as described below.

We note that instead of the mid-plane density and scale height, another set of parameters that
can describe a disk are the disk mass and its spin.
The mid-plane density ($n_0$) is connected to the disk mass ($M_d$) and scale radius (R) via equation \ref{eq:wl}. 
In the scenario of dark matter haloes gaining angular momentum through tidal torque,
it is conventional to use a dimensionless spin parameter $\lambda$ as,
\begin{equation}
 \lambda = J_h |E_h|^{1/2} G^{-1} M_h^{-5/2}\,,
 \label{eq:spin}
\end{equation}
where the  angular momentum of the halo is  $J_h$ and its total energy it $E_h$, with a total mass $M_h$.
The fraction of the halo angular momentum transferred to the disk is parameterised by 
$j_d=J_d/J_h$, where $J_d$ is the disk angular momentum. Thus the scale radius ($R$) is connected to the 
spin parameter via the following equation \citep{mo1998},
\begin{equation}
 R=\bigl({{j_d} \over{\sqrt{2}m_d}}\bigr)\lambda r_{vir}
 \label{eq:scale_radius}
\end{equation}
The scale height is also connected to the mid-plane density via equation \ref{eq:z_distribution}. Therefore,
$n_0$ and $z_0$ are related to the disk mass and the spin parameter.

We recall that 
\citet{wood2000} considered discs with distributed sources, without any dynamical movement of the gas in the disc. For such 
discs, the escape fraction depends only on radiation transfer aspects (and not on opening up of the superbubble), and depend 
on the redshift because of the $n^2$ dependence of the recombination rate. High redshift discs with large gas density would have small 
escape fraction. \citet{wood2000} showed that the escape fraction drops sharply with an increasing redshift because of this.
 In Figure \ref{fig:fesc_z} we show the variation of $f_{\rm esc}$ with the halo mass and 
redshift for $N_{O}=100$ and $10^4$, but without considering the dynamical effects of superbubbles. The case with large $N_{O}$ has large values 
of $f_{\rm esc}$ and falls less sharply with redshift because of a higher ionizing luminosity (see eqn \ref{eq:esc_frac_1_theta}). 
However, when the luminosity function of OB association is taken into account, the low 
$N_{O}$ associations dominate, and therefore the discs in our calculation also show rapid decline of 
$f_{\rm esc}$ with $z$ as in \citet{wood2000}.

We do not expect the \citet{wood2000} prescription for the structure of smooth disks to apply 
at high redshifts because it predicts
extremely large densities and tiny scale heights for even moderate redshifts.
For example, for $M_h=10^{12} M_{\odot}$ at
$z=3$, the density $n_0 \sim 200$ cm$^{-3}$ and a scale height $z_0\sim 13$ pc. 
These disks are likely to cool quickly ($t_{\rm cool}\sim 3.3 \times 10^{-6}$
 Myr), form stars and puff up from feedback effects. These disks will therefore not be able to sustain 
 their high density and small scale heights for long time. Moreover, frequent mergers at high redshifts make 
it difficult for smooth equilibrium discs to survive.

 Observationally, one does not find disc-like structures at redshifts beyond $\sim 2$, and 
high redshift disc galaxies are likely to be clumpy (\citealt{conselice2014}). Drawing from our discussion on 
clumping in the previous section, it is then likely that $f_{\rm esc}$ increases with redshift due to increasing 
clumpiness in discs. Such an increase is consistent with the requirements of reionization (\citealt{mitra2013}).

\begin{figure}
\centerline{
\epsfxsize=0.3\textwidth
\epsfbox{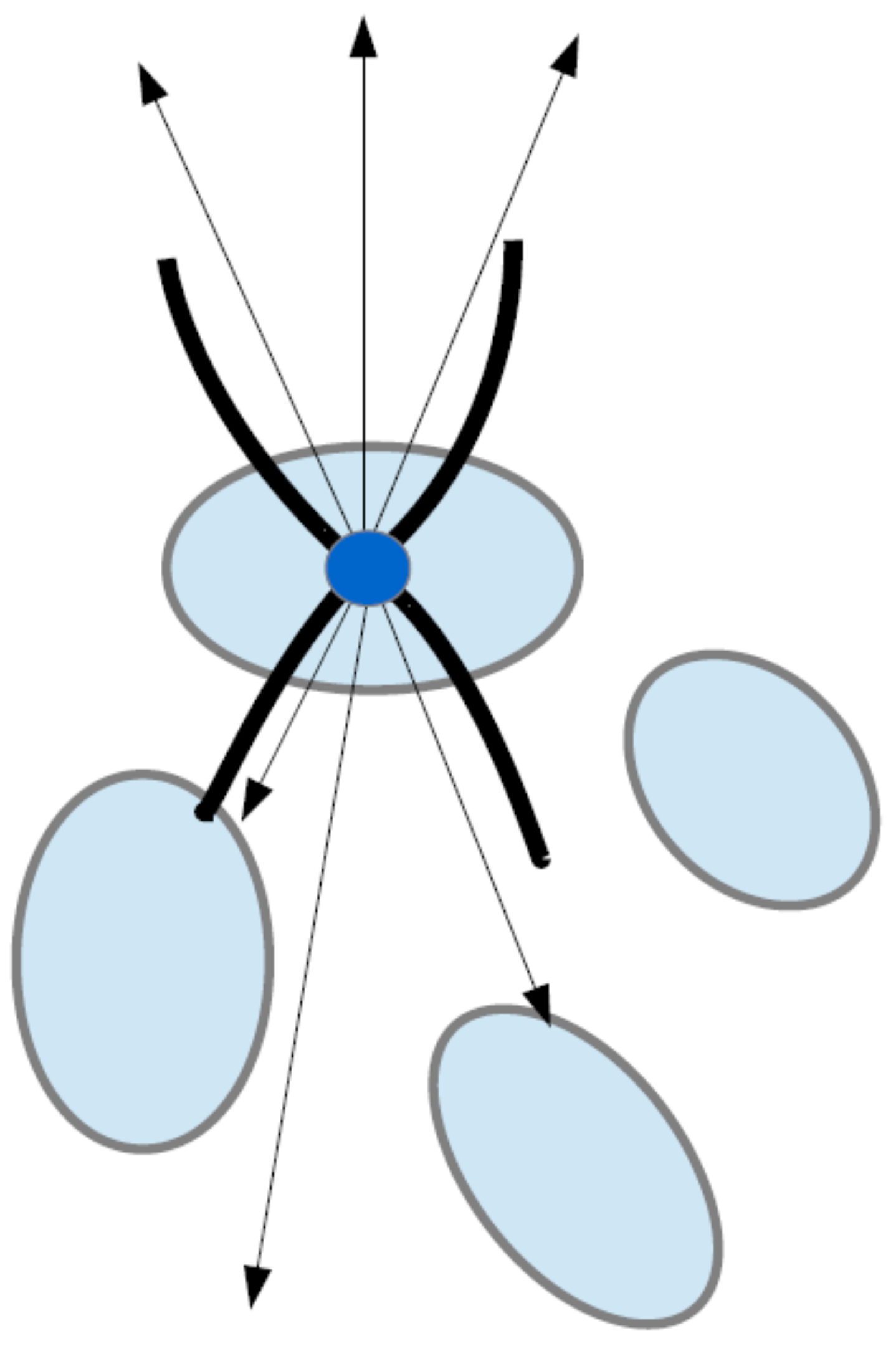}
}
\caption{Schematic diagram of a clumpy ISM at high redshift. The ellipses represent ISM clumps in which star formation occurs; one of them has a
starburst at its center which opens up a superbubble. The thin rays with arrows show LyC photon trajectories through the ISM.
 }
\label{fig:clump_schem_diag}
\end{figure}

Figure \ref{fig:clump_schem_diag} shows a schematic representation of a clumpy ISM expected in a high redshift 
galaxy. We expect the escape fraction to be higher for such a case for several reasons: clumps are more spherical
(the difference between $z_0$ and the scale radius is not as large as in disks) and photon escape both due to radiative 
effects and due to gas dispersal via supernovae can lead to photon escape in all directions (unlike in disks in which 
photons moving along the disk plane are always absorbed); the clumps are themselves more perforated because of
higher star-formation and merger rates at higher redshifts; if most of the clumps are molecular (as is likely in at least in 
star-forming galaxies at high redshifts; \citealt{tacconi2010}) then the escape fraction is higher (as the temperature is 
much lower and since $n_0^2 z_0^3 \propto T$; see eqn \ref{eq:esc_frac_1_theta}). A self-consistent numerical calculation 
is required to calculate the escape fraction in disturbed galaxies 
at high redshifts because of several complexities, but the escape fraction is likely to be larger than in static equilibrium 
disks.

Based on the discussion above, it is then likely that disks at high redshifts did not have extremely large 
density (or very small scale heights) as expected from simple extrapolations of scaling relations. If disks at high
redshifts had gas density in the range $n_0 \le 10$ cm$^{-3}$ and scale height $\ge 100$ pc, then our results
 imply that those disks also had an escape fraction $f_{\rm esc} \sim 10 (\pm 5) \%$, similar to present day disks.
The presence of clumpiness at high redshift would introduce an additional uncertainty, and is likely
to increase the value of  the escape fraction.

 \section{Discussion}
 \label{sec:discussion}
 Our main result is that the escape fraction of ionizing photons depends on the combination of disk 
 gas density and scale height, in particular, on the combination of $n_0^2 z_0^3$. The relation
 between gas density and scale height in turn depends on the disk gas (WNM) temperature.
 The dependence of the escape fraction on other 
 galactic parameters (such as mass and redshift) will be determined by how the WNM temperature and the 
 disk structure depend on these parameters. The  WNM temperature in our Milky Way is known to be $\sim 8000$ K 
 (\citealt{wolfire2003}). It is believed that the WNM temperature increases with decreasing metallicity, 
 because of lack of cooling routes in a low metallicity gas. It is therefore generally 
 thought that the WNM temperature is higher at high redshift, because of metallicity 
 evolution in galaxies. Recent observations of damped Lyman-$\alpha$ systems at $z\sim 3$ support this 
 idea and have found a WNM temperature of $\sim 10^4$ K \citep{cooke2014}. The fact that the spin 
 temperature for 21 cm radiation tends to increase with redshift also supports this 
 \citep{kanekar2014}. Therefore disk galaxies at $z\sim 3$ will have similar escape fractions as in 
 the Milky Way, since the escape
 fraction does not change substantially for WNM temperature in the range of $8000\hbox{--}10^4$ K, 
 as seen in Figure \ref{fig:contour_plot}. At higher redshifts, however, the disk column density is higher
 and the escape fraction is expected to be smaller (see section 4.5 for details).

Our result of an increasing escape fraction with a decreasing galactic mass can be compared with 
previous results. \citet{yajima2011, razoumov2010} have found a decreasing trend of escape fraction
with an increasing galactic mass, although \citet{gnedin2008} found an opposite trend. However, these
results stem from the dependence of clumpiness on disk mass (disks in \citealt{yajima2011}'s simulation
became more clumpy with increasing mass), or the dependence of disk thickness with galactic mass
(in the case of \citealt{gnedin2008}). In contrast, 
our results pertain to smooth disks, and relate the escape fraction in terms of fundamental disk parameters
instead of explicitly coupling it to the galactic mass. We have separately discussed the effect of clumpiness
on the escape fraction, which can be used in conjunction with our basic result.

 Figure \ref{fig:contour_plot} shows that present day disks likely occupy a region of the parameter
 space of density and scale height such that $f_{\rm esc} \sim 10 (\pm 5) \%$ (upper left corner of the
 solid line for isothermal disks). This matches 
 with the requirements for the presence of the Ultraviolet (UV) background radiation in the 
 galaxies \citep{leitherer1995, hurwitz1997, heckman2001}), as well as with direct estimates
 of escape fraction of Milky Way and present day disc galaxies \citep{bland1999, zastrow2013}.
 However, one must bear in mind that, $f_{\rm esc}$ strongly depends on the orientation of the object in the 
plane of the sky, as previously shown \citep{dove2000, fujita2003} and as our calculations confirm. Therefore,
 the observational estimates may be lower than the angle-averaged values presented in this paper.
 
 If high redshift disks have disk gas density $\le 10$ cm$^{-3}$ and scale heights $\ge 100$ pc, then
 our results imply that those disks would also have  $f_{\rm esc} \sim 10 (\pm 5) \%$, modulo the effect
 of clumping. This is
consistent with requirements from  models that explain the epoch of reionization with stellar sources
\citep{madau1996, miralda-escude2000, barkana2001, sommerville2003, robertson2013, gnedin2000,fujita2003, inoue2006,
razoumov2010, yajima2011, paardekooper2013, mitra2013}. Recent numerical simulations by \citet{kimm2014} have found an 
average value of $f_{esc}\sim 10$ \% ($\sim 15$\% including the effect of runaway OB stars) in 
galaxies with mass $10^8\hbox{--}10^{10.5}$ M$_\odot$, and also found this average value to be 
constant over redshift. Our calculations put these results in the perspective of the underlying 
connection between disk parameters and the escape fraction.

\section{Summary \& Conclusions}
\label{sec:conc}
We summarise our conclusions below:
\begin{itemize}
\item Ionizing photons escape from disc galaxies within a cone of angle $\sim 55^\circ$, before the superbubble 
breaks out of the disc, and within  $\sim 40^\circ$ towards the end of the lifetime of OB stars.
\item The ionization cone is unlikely to be larger than an opening angle of $\rm 1$ radian for disk galaxies. This puts an upper limit on the escape fraction for disk galaxies of $0.5$ from geometric considerations.
\item The escape fraction initially decreases, when the superbubble is buried within the disc. It reaches a minimum at 
the breakout epoch and slowly increases with the opening of the shell and the formation of clumps and pathways for ionizing photons.
\item The time-averaged, angle-averaged and OB association luminosity-function-averaged escape fraction for non-clumpy discs depends 
mostly on the WNM temperature and the column density. For typical parameters, we estimate an escape fraction of disk galaxies to be
$\sim 10 (\pm 5) \%$.
\item Escape fraction decreases with the increase of surface covering fraction, although there is no one-to-one correspondence, 
because the escape fraction depends on several factors (such as the gas density, scale height, OB association mass, and time). 
Therefore, the escape fraction for high redshift discs is likely to be higher due to the clumpy nature of ISM at those epochs.
\item A non-negligible escape fraction, especially at high redshifts relevant for reionization, is possible only if superbubbles open 
up a low density ionization cone through which ionizing photons leak into the IGM. The clumpiness of the ISM and the distribution of 
massive stars (which produce most UV photons) play an important role in determining the escape fraction.
\end{itemize}

BBN wishes to thank Andrea Ferrara, Joss Bland-Hawthorn, Sally Oey, Nissim Kanekar and Michael Shull for inspiring discussions. 
BBN and PS acknowledge support from an India-Israel joint research grant (6-10/2014[IC]).

\footnotesize{

\appendix
\section{Convergence test}
We compare the resolutions of $256 \times 128$, $512 \times 256$, and 
$512 \times 512$ for the time and angle dependence of the escape fraction. In Figure \ref{fig:hr_comp_low_n0}
the upper panel shows the time dependence of the $\theta$-averaged escape fraction. Both the subplots are for $n_0=0.5$ cm$^{-3}$, $z_0=300$ pc
and $N_{O}=1000$. 
We notice that initially the escape fraction with low resolution is slightly
higher than with high resolutions. However, after 2-3 $t_{\rm d}$ ($t_{\rm d}=2$ Myr), the escape fraction for 
all  resolutions are similar. 

The bottom panel of Figure \ref{fig:hr_comp_low_n0} shows the angular dependence of the escape fraction for the three resolutions
 at 4 Myr for the same parameters, after the shell fragments due to RTI. One can see the zigzag nature in the high resolution 
 curve, 
 whereas the low resolution curve is comparatively smooth. At higher resolutions,  there are more high density clumps and it results
 in absorption over many angles.  
 
\begin{figure}
\centerline{
\epsfxsize=0.56\textwidth
\epsfbox{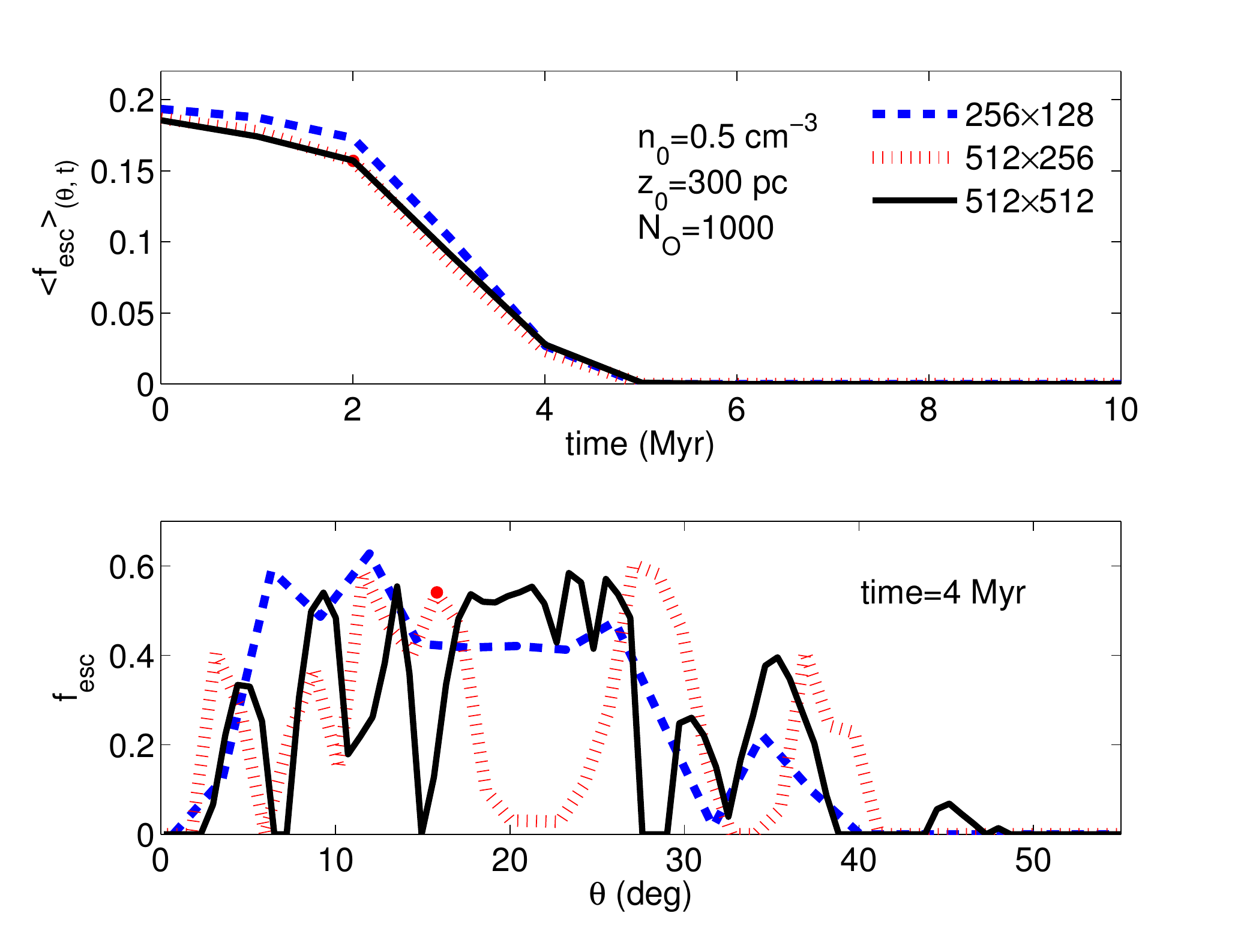}
}
\caption{Comparison between the resolutions of $256 \times 128$, $512 \times 256$ and $512 \times 512$ for the 
angle and time dependence of the escape fraction for $n_0=0.5$ cm$^{-3}$, $z_0=300$ pc, $N_{O}=1000$. The upper 
panel shows the time dependence and the bottom panel represents the angle dependence at 4 Myr, when the fragmentation of 
the shell becomes important due to RTI. All $\langle f_{esc} \rangle_{\theta}$ values in the top panel are zero after 5 Myr.
 }
\label{fig:hr_comp_low_n0}
\end{figure}

In Figure \ref{fig:hr_comp_nob_high_n0}, we show the time dependence of the escape fraction for 
 $256 \times 128$, $512 \times 256$ and $512\times512$ resolution runs for $n_0=15$ cm$^{-3}$, 
 $z_0=30$ pc, and for two different 
 $N_{O}$. The upper and lower panels represent the cases for $N_{O}=300$
 and $N_{O}=10^5$ respectively. The density considered is  high, and consequently there is more clumping at 
 $t<t_d$ for higher resolution. This leads to a high escape fraction in
 high resolutions than that in the low resolution case
 initially.  After the superbubble crosses 
 $2\textendash 3$ scale heights, the different resolution cases start behaving differently 
 due to the different widths of the clumps and channels in them. The bottom panel 
 shows the time dependence of the escape fraction for the same $n_0$ and $z_0$ but for $N_{O}=10^5$. 
 One can notice that the escape fraction at any epoch for all the resolutions are roughly the same and the small changes arise
  from the detailed structure of the shell at a given epoch.

In Figure \ref{fig:hr_comp_low_n0}, the time-averaged and $\theta$-averaged escape fraction for the resolutions of 
$256 \times 128$, $512 \times 256$ and $512 \times 512$ are 
0.1392, 0.1274 and 0.1278 respectively. The percentage change in escape fractions is thus $\le 8$\%. One can also easily notice that at both the 
high resolution ($512 \times 256$ and $512 \times 512$) cases the percentage difference of escape fraction is 0.3\%. In our 
simulations $N_{O}$
 ranges from 100 \citep{zinnecker1993} to $10^5$ (\citealt{ho1997,martin2005,walcher2005})
 and $z_0$ ranges from 10 pc to 600 pc. The maximum difference in the average escape fraction 
 between low ($256 \times 128$) and high resolutions ($512 \times 256$ and $512 \times 512$) for the whole range of 
 $N_{O}$ and $z_0$ used in our simulations for the low $n_0$ ($n_0<5$ cm$^{-3}$) runs is $\approx 10\%$.

 In Figure \ref{fig:hr_comp_nob_high_n0}
  we notice that for low $N_{O}$ ($N_{O}=300$), the average escape fraction for three resolutions ($256 \times 128$, $512 \times 256$ and $512 \times 512$) are 0.023, 0.07 and 0.05 respectively. 
  Thus the percentage change in average escape fraction is 68\% between resolutions of $256 \times 128$ and $512 \times 256$ but  
  it is 28\% between $512 \times 256$ and $512 \times 512$. For 
  high $N_{O}$ case the average escape fraction for low and high resolutions are 0.4165, 0.4243 and 0.4060 respectively, 
  giving a maximum percentage change in escape fraction of
   $\sim 4$\%. We also find that the difference in the average escape fraction between resolutions of 
   $256\times128$ and $512 \times 256$ 
   for large $n_0$ (5 cm$^{-3} \leq n_0 <50$ cm$^{-3}$) cases 
   comes within 10\% for $N_{O} \geq 1000$ for all the values of $z_0$. The high density simulations with $n_0=50$ cm$^{-3}$
   are numerically very expensive and we have to use a relatively low resolution  ($256 \times 128$) for these runs.

    Table \ref{table:res} shows the resolution for our different runs using different $n_0,~z_0$ and $N_O$. We have chosen a
    high enough resolution in each case such that the time- and angle-average escape fraction does not change by more than 20\%
    for a higher resolution run.

\begin{figure}
\centerline{
\epsfxsize=0.56\textwidth
\epsfbox{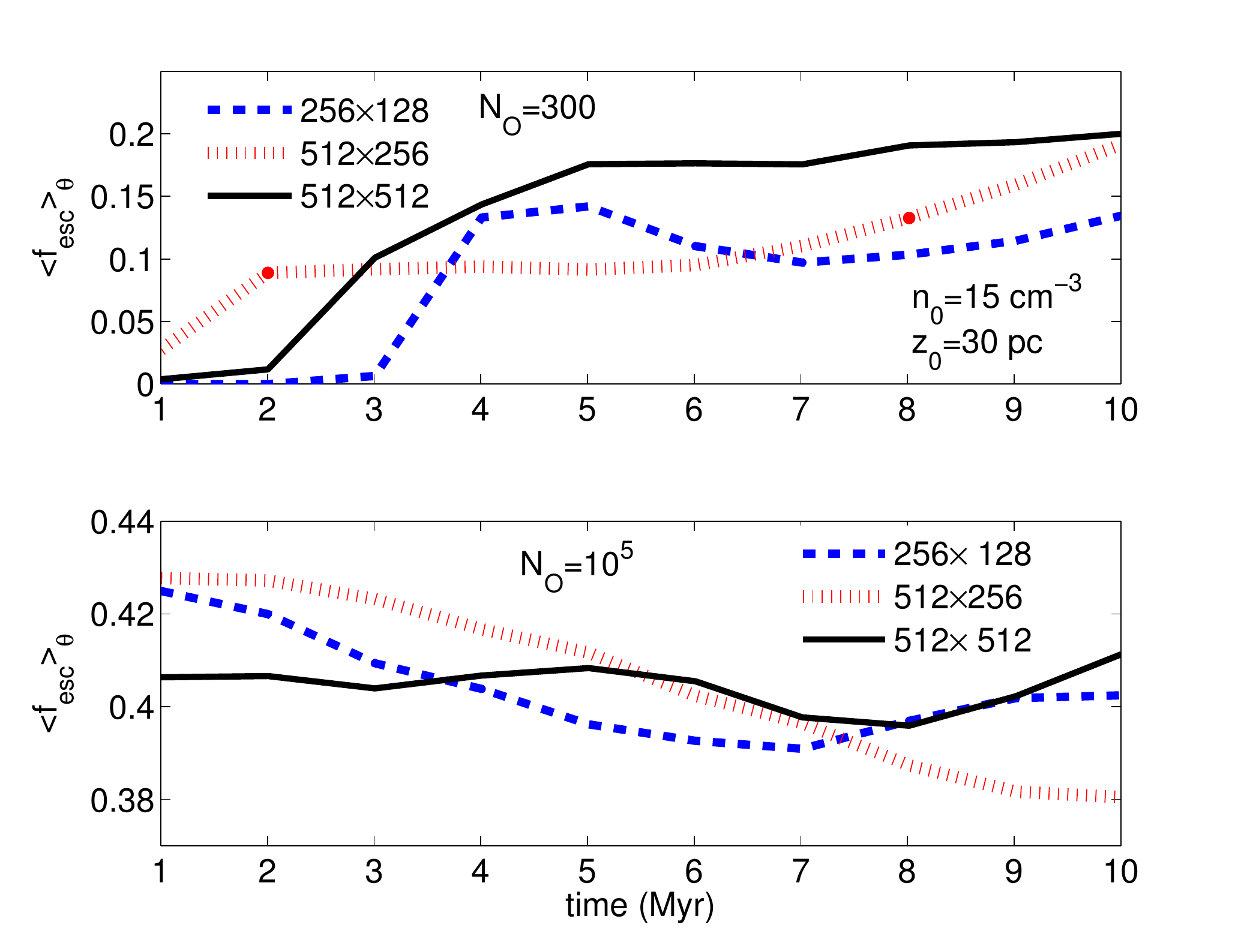}
}
\caption{Comparison between low and high $N_{O}$ ($N_{O}=300$, $10^5$ respectively) cases for the resolutions of $256 \times 
128$, $512 \times 256$ and $512 \times 512$ for the time dependence of escape fraction for 
$n_0=15$ cm$^{-3}$, $z_0=30$ pc.
 }
\label{fig:hr_comp_nob_high_n0}
\end{figure}

\section{Convolution of escape fraction with OB-association}

It is of interest to estimate the value of $N_{O}$ that dominates the process of averaging
$f_{\rm esc}$ over the luminosity function of OB associations. In eqn \ref{eq:final_esc_frac} we have convolved time-averaged and $\theta$-averaged escape fraction with the luminosity function 
of OB-association. On one hand the luminosity function scales as $N_{O}^{-2}$, on the other hand $f_{\rm esc}$ increases with the number of OB stars. Therefore the integrand (of the numerator) 
in eqn \ref{eq:final_esc_frac}  peaks at a certain value of  $N_{O}$.

Figure \ref{fig:lum_fn_nob_plot}  plots the integrand as a function of $N_{O}$ for $n_0=1.5$ cm$^{-3}$ for two different $z_0$ ($60$ pc (the black solid line), $300$
 pc (the red dashed-dotted line)) and $n_0=0.5$ cm$^{-3}$, $z_0=300$ pc (the blue dashed line). For $n_0=1.5$ cm$^{-3}$ and for 
 large scale heights ($300$ pc), the escape fraction decreases to zero for small $N_{O}$ ($\le 1000$) and thus the integrand peaks at higher value of
 $N_{O}$ ($N_{O}=4000$). The integrand decreases with $N_{O}$ for small $z_0$ ($z_0=60$ pc) in the case of $n_0=1.5$ cm$^{-3}$. Thus for the
 small scale heights, the averaging process is dominated by the lowest
 value of $N_{O}$ ($N_{O}=100$). In general, the average escape fraction is dominated by larger $N_O$ for higher $n_0$ and $z_0$.
\begin{figure}
\centerline{
\epsfxsize=0.56\textwidth
\epsfbox{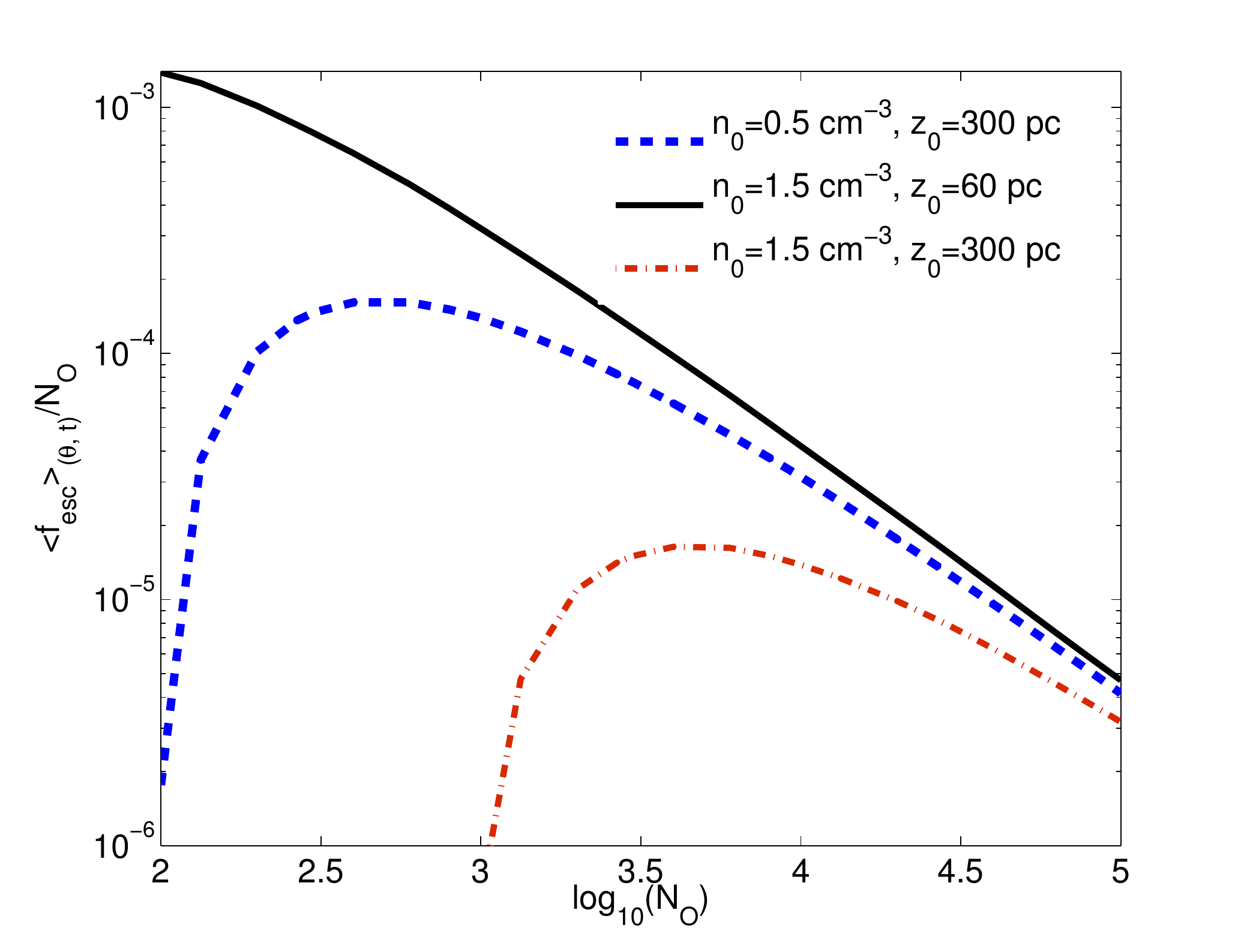}
}
\caption{The integrand of the numerator of eqn  \ref{eq:final_esc_frac} as a function of $N_{O}$ for different $n_0$ and $z_0$.
The blue dashed line represents $n_0=0.5$ cm$^{-3}$, $z_0=300$ pc; the black solid line and the red dashed-dotted 
lines represent $n_0=1.5$ cm$^{-3}$ cases ($z_0=60$ and 300 pc respectively).
}
\label{fig:lum_fn_nob_plot}
\end{figure}

\section{Comparison between 2D and 3D numerical simulations}
\label{app:2D3D}

\begin{figure}
\centerline{
\epsfxsize=0.56\textwidth
\epsfbox{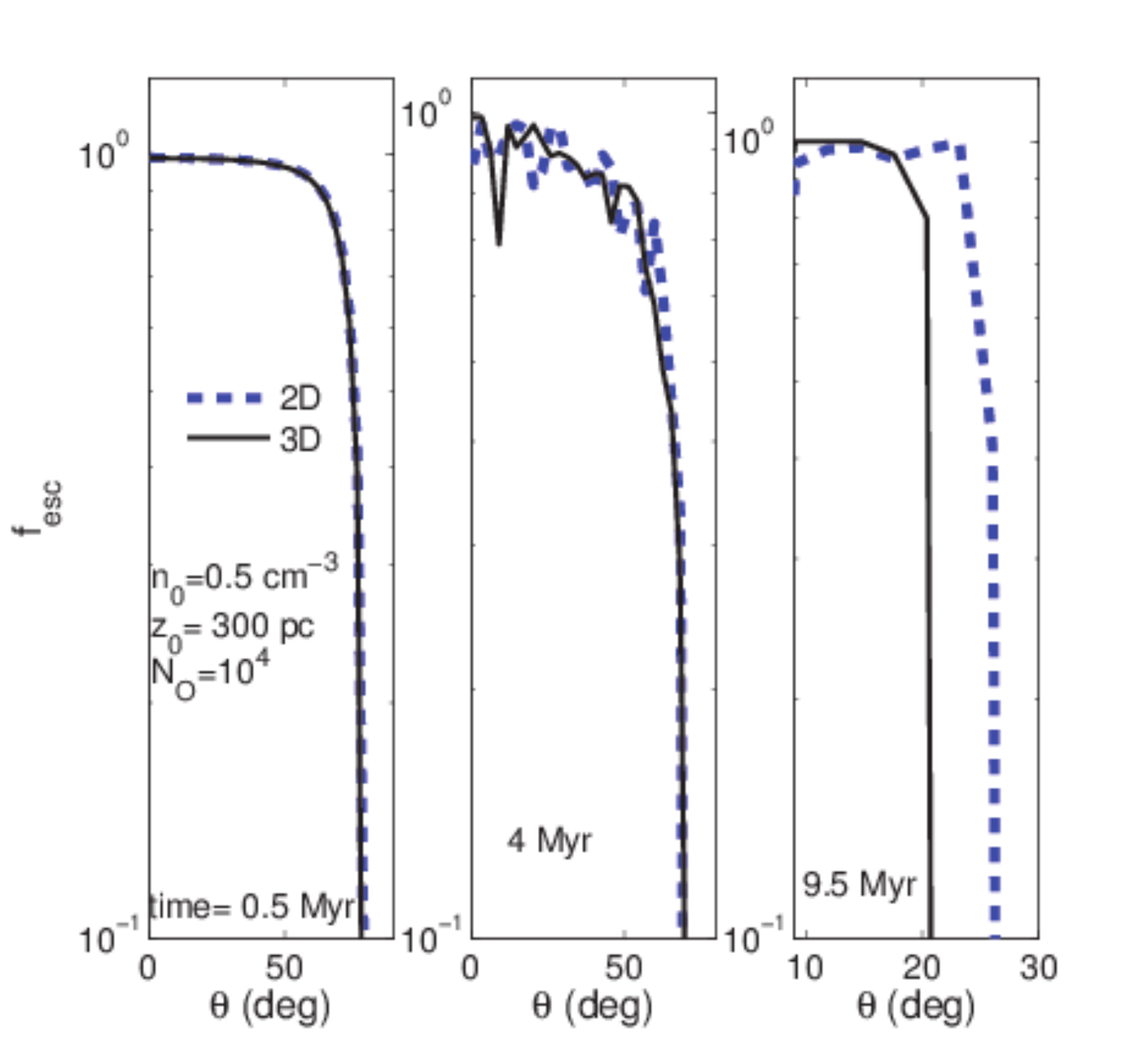}
}
\caption{The comparison of angular variation of the escape fraction between 2D and 3D numerical runs. The blue-dashed and black solid lines 
represent the 2D and 3D runs respectively. The plot is for the fiducial case ($n_0=0.5$ cm$^{-3}$, $z_0=300$ pc, $N_{O}=10^4$).
}
\label{fig:2D_3D_angle}
\end{figure}

\begin{figure}
\centerline{
\epsfxsize=0.56\textwidth
\epsfbox{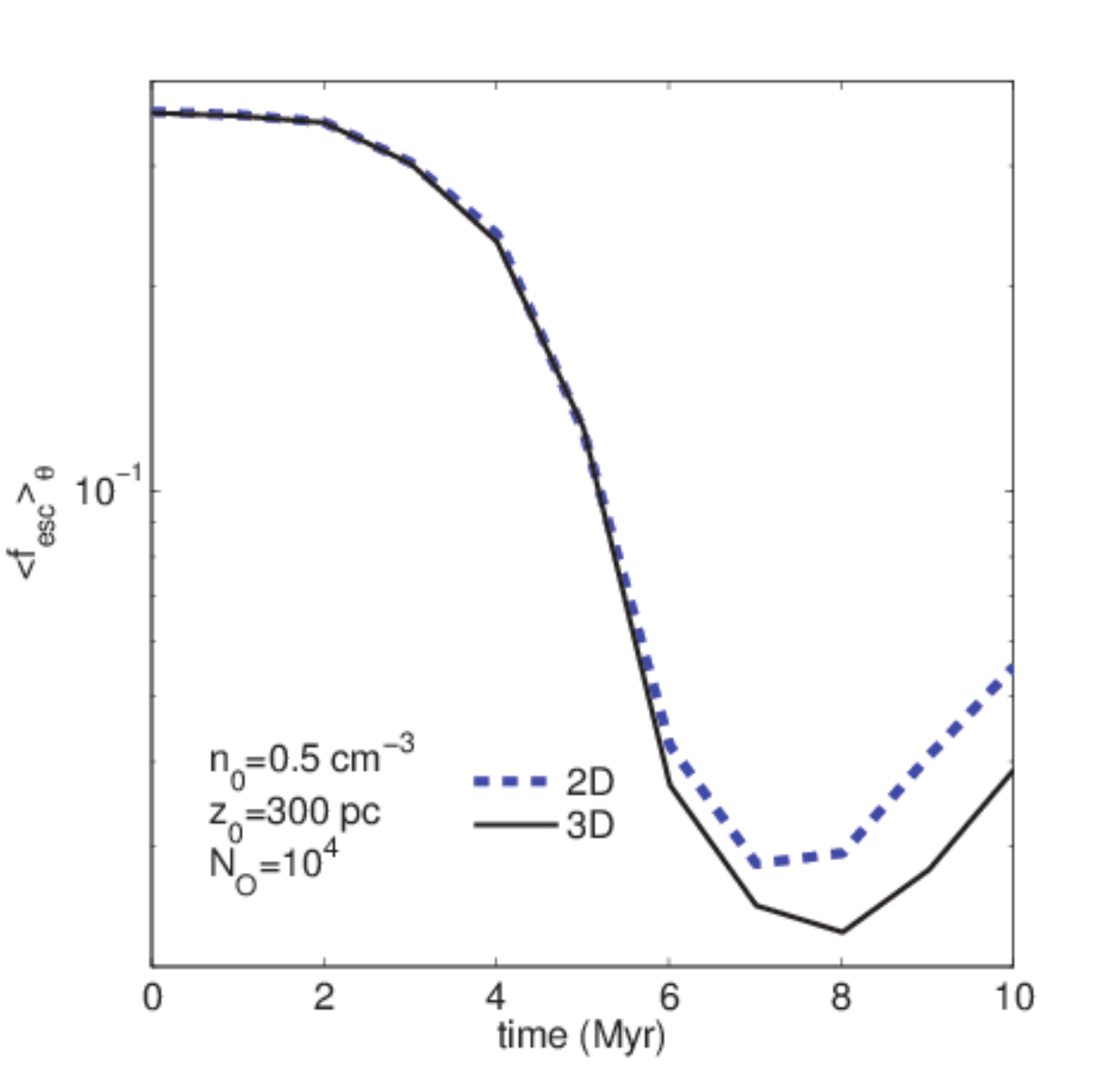}
}
\caption{The comparison of time variation of the escape fraction between 2D and 3D numerical runs. The blue-dashed and black solid lines 
represent the 2D and 3D runs respectively. This plot is also for the fiducial case ($n_0=0.5$ cm$^{-3}$, $z_0=300$ pc, $N_{O}=10^4$).
}
\label{fig:2D_3D_time}
\end{figure}
In this section we show the angle and time variation of the escape fraction for 2D ($256 \times 128$) and 3D ($256 \times 128 \times 32$) simulations 
for the fiducial case ($n_0=0.5$ cm$^{-3}$, $z_0=300$ pc, $N_{O}=10^4$).
In 3D simulation we have uniformly spaced grid points in the $\phi$-direction.

Figure \ref{fig:2D_3D_angle} shows the angular variation of the escape fraction for the 2D and 3D runs at three different times.
The angular variation of the escape fraction in 2D and 3D matches at early times (0.5 Myr). 
 At a later epoch (4.0 Myr), when RTI starts playing a crucial role 
for the detailed structure of the shell, the angular variations show some 
differences.
The differences are small enough to have a negligible effect on the final $<f_{esc}>_{\theta}$ 
(refer to the figure \ref{fig:2D_3D_time}). At very late times the escape cone is slightly larger in 2D compared to 3D, which makes $<f_{esc}>_{\theta}$
a bit higher in 2D. The time-averaged theta-averaged escape fraction in 2D and 3D (0.315 and 0.3137, respectively) are very similar. Thus, the
use of the faster 2D simulations to calculate the escape fraction is justified.


\begin{thebibliography}{}

\bibitem[\protect\citeauthoryear{Adelberger \& Steidel}{2000}]{adelberger2000}
Adelberger, K. L., Steidel, C. C. 2000, ApJ, 544, 218

\bibitem[\protect\citeauthoryear{Altay \etal}{2011}]{altay2011}
Altay, G., Theuns, T., Schaye, J., Vrighton, N. H. M., Dalla Vecchia, C. 2011, ApJ, 737, L37

\bibitem[\protect\citeauthoryear{Barkana \& Loeb}{2001}]{barkana2001}
Barkana, R., Loeb, A. 2001, Phys. Rep., 349, 125

\bibitem[\protect\citeauthoryear{Benson \etal}{2013}]{benson2013}
Benson, A., Venkatesan, A. Shull, J. M. 2013, ApJ, 770, 76

\bibitem[\protect\citeauthoryear{Bland-Hawthorn \& Maloney}{1999}]{bland1999}
Bland-Hawthorn, J., Maloney, P. R. 1999, ApJL, 510, 33

\bibitem[\protect\citeauthoryear{Borthakur \etal}{2014}]{borthakur2014}
Borthakur, S., Heckman, T. M., Leitherer, C., Overzier, R. A. 2014, Science, 346, 216

\bibitem[\protect\citeauthoryear{Cen}{2012}]{cen2012}
Cen, R. 2012, ApJ, 748, 121

\bibitem[\protect\citeauthoryear{Chiosi, Nasi \& Sreenivasan}{1978}]{chiosi1978}
Chiosi, C., Nasi, E., Sreenivasan, S. P. 1978, Astr. Ap., 63, 103

\bibitem[\protect\citeauthoryear{Clarke \& Oey}{2002}]{clarke2002}
Clarke, C., Oey, M. S. 2002, MNRAS, 337, 1299

\bibitem[\protect\citeauthoryear{Conselice}{2014}]{conselice2014}
Conselice, C. 2014, arxiv:1403.2783

\bibitem[\protect\citeauthoryear{Cooke \etal}{2014}]{cookej2014}
Cooke, J., Ryan-Weber, E. V., Garel, T., D\'iaz, C. G. 2014, MNRAS, 441, 837

\bibitem[\protect\citeauthoryear{Cooke \etal}{2014}]{cooke2014}
Cooke, R. J., Pettini, M., Jorgenson, R. A., 2014, arXiv:  1406.7003

\bibitem[\protect\citeauthoryear{Dove \& Shull}{1994}]{dove1994}
Dove, J. B., Shull, J. M. 1994, ApJ, 430, 222

\bibitem[\protect\citeauthoryear{Dove \etal}{2000}]{dove2000}
Dove, J. B., Shull, J. M. , Ferrara, A. 2000, ApJ, 531, 846

\bibitem[\protect\citeauthoryear{Dyson \& Williams}{1997}]{dyson1997}
Dyson, J. E., Williams, D. A. 1997, The Physics of the Interstellar Medium, Second Edition

\bibitem[\protect\citeauthoryear{Erkal \etal}{2012}]{erkal2012}
Erkal, D., Gnedin, N. Y. , Kravtsov, A. V. 2012, ApJ, 761, 54

\bibitem[\protect\citeauthoryear{Fernandez \& Shull }{2011}]{fernandez2011}
Fernandez, E. R., Shull, J. M. 2011, ApJ, 731, 20

\bibitem[\protect\citeauthoryear{Ferrara \& Loeb }{2013}]{ferrara2013}
Ferrara, A., Loeb, A. 2013, MNRAS, 431, 2826

\bibitem[\protect\citeauthoryear{Fujita \etal}{2003}]{fujita2003}
Fujita, A., Martin, C., MacLow, M., Abel, T. 2003, ApJ, 599, 50

\bibitem[\protect\citeauthoryear{Fumagalli \etal}{2011}]{fumagalli2011}
Fumagalli, M., Prochaska, J. X., Kasen, D., Dekel, A., Ceverino, D., Primack, J. R. 2011, MNRAS, 418, 1796

\bibitem[\protect\citeauthoryear{Gnedin}{2000}]{gnedin2000}
Gnedin, N. 2000, ApJ, 535, 530

\bibitem[\protect\citeauthoryear{Gnedin \etal}{2008}]{gnedin2008}
Gnedin, N. Y., Kravtsov, A. V., Chen, H.-W. 2008, ApJ, 672, 765

\bibitem[\protect\citeauthoryear{Haardt \& Madau}{2012}]{haardt2012}
Haardt, F., Madau, P. 2012, ApJ, 746, 125

\bibitem[\protect\citeauthoryear{Haehnelt \etal}{2001}]{haehnelt2001}
Haehnelt, M. G., Madau, P., Kudritzki, R., Haardt, F. 2001, ApJ, 549, 151

\bibitem[\protect\citeauthoryear{Hayes \etal}{2006}]{hayes2006}
Hayes, J. C., Norman, M. L., Fiedler, R. A., Bordner, J. O., Li, P. S., Clark, S. E., ud-Doula, A., Mac Low, M-M 2006, ApJS, 165, 188


\bibitem[\protect\citeauthoryear{Heckman \etal}{2001}]{heckman2001}
Heckman, T., Sembach, K., Meurer, G., Leithere, C., Calzetti, D., Martin, C. 2001, ApJ, 558, 56

\bibitem[\protect\citeauthoryear{Heckman \etal}{2011}]{heckman2011}
Heckman, T., Borthakur, S., Overzier, R. \etal 2011, ApJ, 730, 5

\bibitem[\protect\citeauthoryear{Hirashita \& Ferrara}{2005}]{hirashita2005}
Hirashita, H., Ferrara, A. 2005, MNRAS, 356, 1529

\bibitem[\protect\citeauthoryear{Ho}{1997}]{ho1997}
Ho, P. 1997, Rev. Mex AA Conference Ser., 6, 5

\bibitem[\protect\citeauthoryear{Hurwitz \etal}{1997}]{hurwitz1997}
Hurwitz, M., Jelinsky, P., Dixon, W. V. 1997, ApJ, 481, L31

\bibitem[\protect\citeauthoryear{Hutter \etal}{2014}]{hutter2014}
Hutter, A., Dayal, P., Partl, A. M., M\"uller, V. 2014, MNRAS, 441, 2861


\bibitem[\protect\citeauthoryear{Inoue \etal}{2006}]{inoue2006}
Inoue, A. K., Iwata, I., Deharveng, J.-M., 2006, MNRAS, 371, L1

\bibitem[\protect\citeauthoryear{Kanekar \etal}{2011}]{kanekar2011}
Kanekar, N., Braun, R., Roy, N. 2011, ApJ, 737, L33

\bibitem[\protect\citeauthoryear{Kanekar \etal}{2014}]{kanekar2014}
Kanekar, N., Prochaska, J. X., Smette, A. \etal 2014, MNRAS, 438, 2131

\bibitem[\protect\citeauthoryear{Kimm \& Cen}{2014}]{kimm2014}
Kimm, T. \& Cen, R. 2014, ApJ, 788, 121

\bibitem[\protect\citeauthoryear{Kollmeier \etal}{2014}]{kollmeier2014}
Kollmeier, J. A., Weinberg, D. H., Oppenheimer, B. D. \etal 2014, ApJL, 789, L32

\bibitem[\protect\citeauthoryear{Kompaneets}{1960}]{kompaneets1960} 
 Kompaneets, A. S. 1960, Soviet Phys Dokl., 5. 46
 
 \bibitem[\protect\citeauthoryear{Koyama \& Inutsuka}{2004}]{koyama2004} 
 Koyama, H. \& Inutsuka, S. 2004, ApJ, 602, L25
 
 \bibitem[\protect\citeauthoryear{Leitet \etal}{ 2013}]{leitet2013}
 Leitet, E., Bergvall, N., Hayes, M., Linn\'e, Zackrisson, E. 2013, A\&A, 553, 106L

\bibitem[\protect\citeauthoryear{Leitherer \etal}{1995}]{leitherer1995}
Leitherer, C., Ferguson, H. C., Heckman, T. M, Lowenthal, J. D. 1995, ApJ, 454, L19

\bibitem[\protect\citeauthoryear{Leitherer \etal}{1999}]{leitherer1999} 
Leitherer C. \etal, 1999, ApJS, 123, 3 


\bibitem[\protect\citeauthoryear{Madau \& Shull}{1996}]{madau1996}
Madau, P., Shull, J. M., 1996, ApJ, 457, 551

\bibitem[\protect\citeauthoryear{Madau \etal}{1999}]{madau1999}
Madau, P., Haardt, F. \& Rees, M. J., 1999, ApJ, 514, 648

\bibitem[\protect\citeauthoryear{Mart\'{i}n \etal}{2005}]{martin2005}
Mart\'{i}n-Hern\'{a}ndez, N. L., Schaerer, D., Sauvage, M. 2005, A\&A, 429, 449

\bibitem[\protect\citeauthoryear{McKee \& Williams}{1997}]{mckee1997}
McKee, C. F., Williams, J. P. 1997, ApJ, 476, 144

\bibitem[\protect\citeauthoryear{Meynet \& Maeder}{2003}]{meynet2003}
Meynet, G., Maeder, A. 2003, A\&A, 404, 975

\bibitem[\protect\citeauthoryear{Miralda-Escud\'e \etal}{2000}]{miralda-escude2000}
Miralda-Escud\'e, J., Haehnelt, M., Rees, M. J.  2000, ApJ, 530, 1


\bibitem[\protect\citeauthoryear{Mitra \etal}{2013}]{mitra2013}
Mitra, S., Ferara, A., Choudhury, T. R., 2013, MNRAS, 428, L1

\bibitem[\protect\citeauthoryear{Mo, Mao and White}{1998}]{mo1998}
Mo H. J., Mao S., White S. D. M., 1998, MNRAS, 295, 319

\bibitem[\protect\citeauthoryear{Navarro \etal} {1997}]{nfw1997}
Navarro J. F., Frenk C. S., White S. D. M., 1997, ApJ, 490, 493

\bibitem[\protect\citeauthoryear{Nestor \etal} {2011}]{nestor2011}
Nestor, D. B., Shapley, A. E., Steidel, C. C., Siana, B. 2011, ApJ, 736, 18

\bibitem[\protect\citeauthoryear{Paardekooper \etal}{2013}]{paardekooper2013}
Paardekooper, J.-P., Khochfar, S., Dalla Vecchia, C. 2013, MNRAS, 429, L94

\bibitem[\protect\citeauthoryear{Pellegrini \etal}{2012}]{pellegrini2012}
Pellegrini, E. W., Oey, M. S., Winkler, P. F., Points, S. D., Smith, R. C., Jaskot, A. E., Zastrow, J.
2012, ApJ, 755, 40


\bibitem[\protect\citeauthoryear{Prochaska \etal} {2005}]{prochaska2005}
Prochaska, J. X., Herbert-Fort, S., Wolfe, A. M. 2005, ApJ, 635, 123

\bibitem[\protect\citeauthoryear{Razoumov \& Sommer-Larsen}{2010}]{razoumov2010}
Razoumov, A. O, Sommer-Larsen, J. 2010, ApJ, 710, 1239

\bibitem[\protect\citeauthoryear{Robertson \etal}{2013}]{robertson2013}
Robertson, B. E., Furlanetto, S. R., Schneider, E. \etal 2013, ApJ, 768, 71

\bibitem[\protect\citeauthoryear{Roy \etal}{2013}]{roy2013}
Roy A., Nath B. B., Sharma P., Shchekinov Y., 2013, MNRAS, 434, 3572

\bibitem[\protect\citeauthoryear{Schaye}{2001}]{schaye2001}
Schaye, J. 2001, ApJ, 562, L95

\bibitem[\protect\citeauthoryear{Sharma \etal}{ 2014}]{sharma2014}
Sharma, P., Roy, A., Nath, B. B., Shchekinov Y., 2014, MNRAS, 443, 3463

\bibitem[\protect\citeauthoryear{Shull \etal}{ 1999}]{shull1999}
Shull, J. M., Roberts, D., Giroux, M. L., Penton, S. V., Fardal, M. A. 1999, AJ, 118, 1450

\bibitem[\protect\citeauthoryear{Siana \etal}{2010}]{siana2010}
Siana, B., Teplitz, H. I., Ferguson, H. C. 2010, ApJ, 723, 241

\bibitem[\protect\citeauthoryear{Sommerville \etal}{2003}]{sommerville2003}
Sommerville, R. S., Bullock, J. S., Livio, M. 2003, ApJ, 593, 616

\bibitem[\protect\citeauthoryear{Sutherland \& Dopita}{1993}]{sutherland1993}
Sutherland, R. S., Dopita, M. A. 1993, ApJS, 88, 253

\bibitem[\protect\citeauthoryear{Tacconi \etal}{2010}]{tacconi2010}
Tacconi, L. J. et al. 2010, Nature, 463, 781

\bibitem[\protect\citeauthoryear{Vacca \etal} {1996}]{vacca1996}
Vacca, W. D., Garmany, C. D., Shull, J. M., 1996, ApJ, 460, 914

\bibitem[\protect\citeauthoryear{Walcher \etal} {2005}]{walcher2005}
Walcher, C. J., et al. 2005, ApJ, 618, 237

\bibitem[\protect\citeauthoryear{Weaver \etal} {1977}]{weaver1977}
Weaver R., McCray R., Castor J., Shapiro P., Moore R., 1977, ApJ, 218, 377

\bibitem[\protect\citeauthoryear{Weidner \etal} {2010}]{weidner2010}
Weidner C., Vink J. S. 2010, A\&A, WVostars Rev.

\bibitem[\protect\citeauthoryear{Wise \etal} {2014}]{wise2014}
Wise, J. H., Demchenko, V. G., Halicek, M. T., Norman, M. L., Turk, M. J.,Abel, T., Smith, B. D. 2014, MNRAS, 442, 2560

\bibitem[\protect\citeauthoryear{Wolfire \etal} {2003}]{wolfire2003}
Wolfire, M. G., McKee, C. F., Hollenbach, D., Tielens, A. G. G. M. 2003, ApJ, 587, 278

\bibitem[\protect\citeauthoryear{Wood \& Loeb} {2000}]{wood2000}
Wood, K., Loeb, A. 2000, ApJ, 545, 86

\bibitem[\protect\citeauthoryear{Yajima \etal}{ 2011}]{yajima2011}
Yajima, H., Choi J.-H., Nagamine, K. 2011, MNRAS, 412, 411


\bibitem[\protect\citeauthoryear{Zastrow \etal}{2013}]{zastrow2013}
Zastrow, J., Oey, M. S., Veilleux, S., McDonald, M. 2013, ApJ, 779, 76

\bibitem[\protect\citeauthoryear{Zinnecker \etal}{1993}]{zinnecker1993}
Zinnecker, H., McCaughrean, M. J., Wilking, B. A. 1993, in Protostars and Planets III, ed. 
E. H. Levy \& J. I. Lunine (Univ. of Arizona Press), 429

\bibitem[\protect\citeauthoryear{Zwaan \& Prochaska}{2006}]{zwaan2006}
Zwaan, M. A., Prochaska, J. X. 2006, ApJ, 643, 675

\bibitem[\protect\citeauthoryear{Zwaan \etal}{2005}]{zwaan2005}
Zwaan, M. A., van der Hulst, J. M., Briggs, F. H., Verheijen, M. A. W. 2005, MNRAS, 364, 1467
\end{thebibliography}
\end{document}